\documentclass[a4paper,twocolumn,11pt,accepted=2025-03-20]{quantumarticle}
\pdfoutput=1
\usepackage[utf8]{inputenc}
\usepackage[english]{babel}
\usepackage[T1]{fontenc}
\usepackage{amsmath}
\usepackage{hyperref}
\usepackage{bm}
\usepackage[numbers,sort&compress]{natbib}
\usepackage{tikz}
\usepackage{lipsum}

\usepackage{amsmath,amssymb,amsfonts,float,graphics,epsfig,epstopdf,color,verbatim,tabularx,bm,multirow,appendix}
\usepackage{xcolor,mathtools}
\usepackage{dsfont}
\usepackage{textcomp}
\usepackage{footnote}
\usepackage{bm}
\usepackage{subfigure}
\usepackage{mathrsfs,enumitem}
\usepackage{graphicx}
\usepackage{verbatim}
\usepackage{multirow}
\usepackage{braket}
\usepackage[normalem]{ulem}
\usepackage{multirow}
\usepackage{makecell}
\usepackage{tikz}
\usepackage{lipsum}
\usetikzlibrary{calc}
\usetikzlibrary{shapes.multipart}

\usepackage{environ}

\NewEnviron{myequation}{%
    \begin{equation}
    \scalebox{0.87}{$\displaystyle{\BODY}$}
    \end{equation}
    }

\newcommand{\comments}[1]{}

\newcommand{\av}[1]{\langle#1\rangle}

\def\renyione{\overline{S^{(2)}_{A,1}}}

\def\aone{a^{(2)}_1}
\def\atwo{a^{(2)}_2}
\def\athree{a^{(2)}_3}

\newcommand{\stkout}[1]{\ifmmode\text{\sout{\ensuremath{#1}}}\else\sout{#1}\fi}

\makeatletter
\def\l@subsubsection#1#2{}
\makeatother

\begin{document}

\title{Monte Carlo Simulation of Operator Dynamics and Entanglement in Dual-Unitary Circuits}

\author{Menghan Song}
\affiliation{Department of Physics and HK Institute of Quantum Science \& Technology, The University of Hong Kong, Pokfulam Road, Hong Kong}

\author{Zhao-Yi Zeng}
\affiliation{Department of Physics, Fudan University, Shanghai, 200438, China}

\author{Ting-Tung Wang}
\affiliation{Department of Physics and HK Institute of Quantum Science \& Technology, The University of Hong Kong, Pokfulam Road, Hong Kong}

\author{Yi-Zhuang You}
\email{yzyou@physics.ucsd.edu}
\affiliation{Department of Physics, University of California, San Diego, California 92093, USA}

\author{Zi Yang Meng}
\email{zymeng@hku.hk}
\affiliation{Department of Physics and HK Institute of Quantum Science \& Technology, The University of Hong Kong, Pokfulam Road, Hong Kong}

\author{Pengfei Zhang}
\email{pengfeizhang.physics@gmail.com}
\affiliation{Department of Physics, Fudan University, Shanghai, 200438, China}
\affiliation{State Key Laboratory of Surface Physics, Fudan University, Shanghai, 200438, China}
\affiliation{Shanghai Qi Zhi Institute, AI Tower, Xuhui District, Shanghai 200232, China}
\affiliation{Hefei National Laboratory, Hefei 230088, China}

\maketitle

\begin{abstract}
		We investigate operator dynamics and entanglement growth in dual-unitary circuits, a class of locally scrambled quantum systems that enables efficient simulation beyond the exponential complexity of the Hilbert space. By mapping the operator evolution to a classical Markov process, we perform Monte Carlo simulations to access the time evolution of local operator density and entanglement with polynomial computational cost. Our results reveal that the operator density converges exponentially to a steady-state value, with analytical bounds that match our simulations. Additionally, we observe a volume-law scaling of operator entanglement across different subregions, and identify a critical transition from maximal to sub-maximal entanglement growth, governed by the circuit’s gate parameter. This transition, confirmed by both mean-field theory and Monte Carlo simulations, provides new insights into operator entanglement dynamics in quantum many-body systems. Our work offers a scalable computational framework for studying long-time operator evolution and entanglement, paving the way for deeper exploration of quantum information dynamics.
\end{abstract}

\section{Introduction}
\label{sec:intro}

In quantum many-body physics, the study of operator dynamics is an active area of research. The dynamics of quantum operators under Heisenberg picture time evolution can provide insights into the complex behavior of quantum many-body systems, including operator spreading \cite{Ho2017E1508.03784,Nahum2017Q1608.06950,Nahum2018O1705.08975,von-Keyserlingk2018O1705.08910,Khemani2018O1710.09835,Chan2018S1712.06836,Zhou2019O1805.09307,Qi2019M1906.00524,von-Keyserlingk2022O2111.09904,Schuster2022O2208.12272}, scrambling and thermalization \cite{Bohrdt2017S1612.02434,Kukuljan2017W1701.09147,Xu2019L1805.05376,Parker2019A1812.08657,Kuo2020M1910.11351,Akhtar2020M2006.08797,Zhang2023U2305.02356,Liu2023K2207.13603,Buca2023U2301.07091,Zhang:2022fma,Zhang:2022atf}. Simulating operator dynamics in quantum many-body systems is generally challenging due to the curse of dimensionality—the exponential growth of the Hilbert space with system size. While efficient computational methods exist in specific cases, such as Clifford circuits~\cite{Gottesman1998Tquant-ph/9807006}, time-evolving block decimation (TEBD)~\cite{Vidal2004Equant-ph/0310089,haegemanTime2011,haegemanUnifying2016}, or systems with partial integrability~\cite{muthDynamical2011,albaOperator2019}, a general solution remains elusive.

When focusing solely on the statistical properties of \emph{operator dynamics}, i.e.~how the size and support of an operator evolves on average without concern for the specific operator content, more efficient approaches become available~\cite{Nahum2018O1705.08975,jonayCoarse2018,von-Keyserlingk2018O1705.08910,Zhou2019O1805.09307,Zhou2019E1804.09737,chen2024freeindependencenoncrossingpartition}. For \emph{locally scrambled} quantum dynamics~\cite{Kuo2020M1910.11351,Akhtar2020M2006.08797}, where the unitary evolution operators are weakly symmetric~\cite{de-Groot2022S2112.04483,Ma2023A2209.02723} under local basis transformations, the operator dynamics can be mapped to an effective classical diffusion model in the space of operator supports~\cite{Nahum2018O1705.08975,Zhou2019E1804.09737,Nahum2022R2205.11544}. This mapping simplifies the quantum evolution into a tractable classical Markov process, allowing for efficient simulation via Monte Carlo (MC) sampling. 

Furthermore, to characterize the internal complexity and correlations within an operator, the concept of \emph{operator entanglement} has been introduced~\cite{Nie2019S1812.00013,Kudler-Flam2020Q1906.07639,bertiniOperatorII2020,bertiniOperatorI2020}. Operator entanglement quantifies the entanglement of the corresponding Choi state representation~\cite{Jamiokowski1972L,Choi1975C} of the operator in the operator Hilbert space (a.k.a.~doubled Hilbert space). In the context of operator dynamics, it characterizes how the information encoded by the initial operator scrambles with time and becomes correlated across different parts of the system over time. Nicely, we found that the second R\'enyi operator entanglement can also be studied under reasonable assumptions using above-mentioned Monte Carlo approach for operator dynamics on multiple replicas, enabling efficient computation of these quantities in complex systems.

In this work, we investigate the operator dynamics and operator entanglement in \emph{dual-unitary circuits} --- a class of quantum circuits consisting of dual-unitary gates, which are unitary in both space and time directions~\cite{bertiniExact2019,bertiniOperatorI2020,bertiniOperatorII2020}. Dual-unitary circuits, like Haar random unitary circuits, are examples of locally scrambled quantum dynamics that enable operator dynamics to be studied efficiently by mapping to classical diffusion processes. However, compared to Haar random unitary gates, dual-unitary gates form a family of random unitary ensembles with tunable scrambling power controlled by a single parameter. This tunability allows us to systematically explore different scrambling behaviors within the same theoretical framework.


We demonstrate, via Monte Carlo simulations of the local operator density and entanglement in a dual-unitary circuit, that the operator density near the light cone converges exponentially to $\frac{3}{4}$, with a rate determined by the gate parameter  $\alpha$. In the long-time limit, we derive analytical bounds for the density profile, validated by Monte Carlo simulations. For operator entanglement, we observe volume-law scaling across different subregions, with the volume-law coefficient’s dependence on $\alpha$ varying. A key finding is the transition from maximal to sub-maximal entanglement growth for left movers at the left light cone, consistent with prior theoretical work~\cite{bertiniExact2019,bertiniOperatorI2020,bertiniOperatorII2020,akhtar2024dual}. We provide a concrete physical explanation for the emergence of the entanglement transition based on operator emission, and clarify the setup for observing the transition in the physical regime by focusing on left-movers only. Furthermore, our study provides a direct mapping of the volume-law coefficient to $\alpha$, confirmed by mean-field (MF) theory and Monte Carlo simulations. Our results offering new insights into operator dynamics and entanglement transitions in dual-unitary circuits.

\begin{figure}[t!]
\includegraphics[width=\columnwidth]{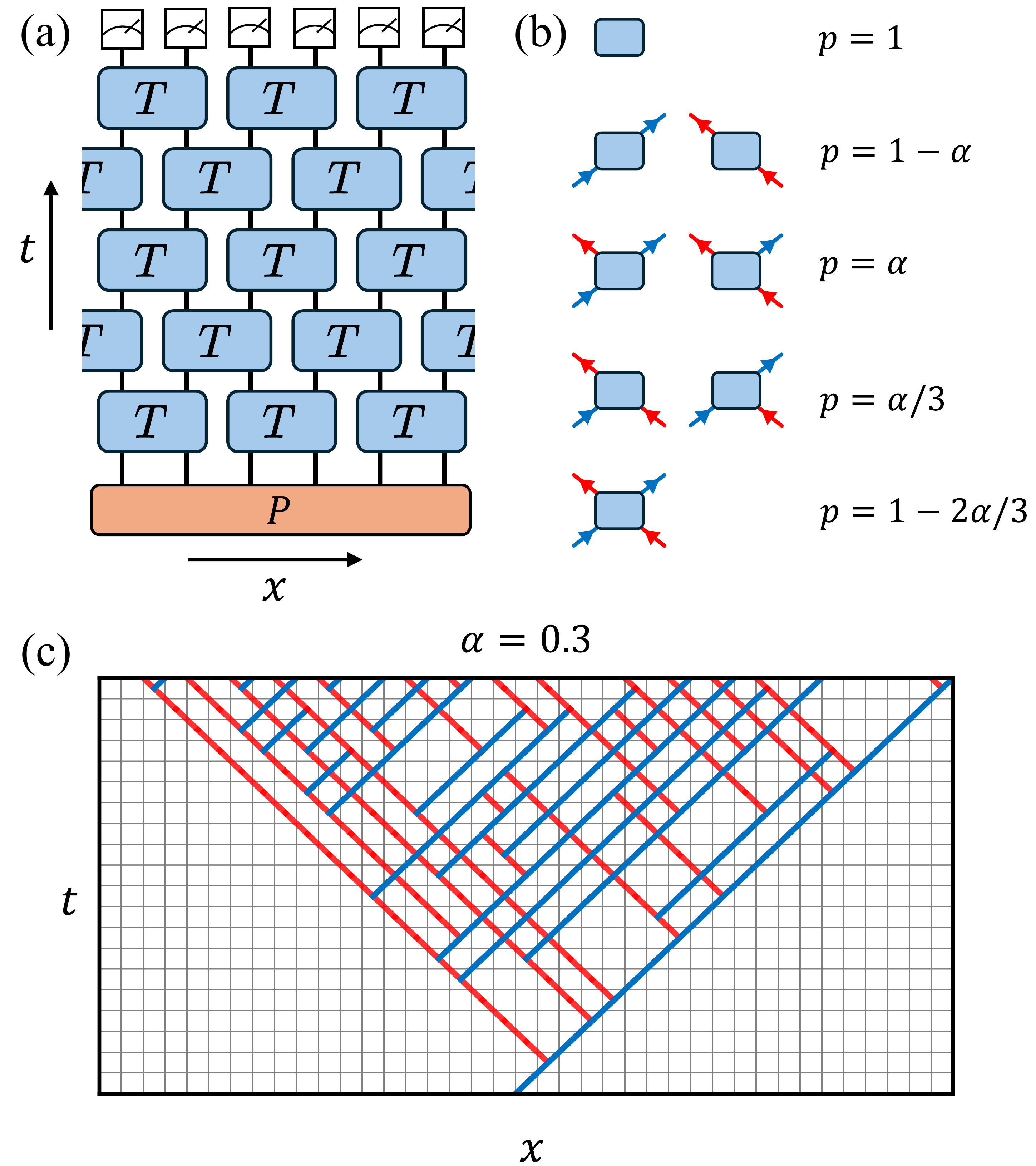}
\caption{\textbf{Schematic plot of a dual-unitary circuit which evaluates the Pauli weight $w(P)$ on a Pauli string $P$.} (a) The brick-wall structure of local dual-unitary gates as transfer matrices $T$. The circuit takes a Pauli string $P=\otimes_i P_i$ as an input, where local Pauli matrix $P_i\in\{I, X, Y, Z\}$. Operators $\{X, Y, Z\}$ and $I$ are capsuled as $\frac{1}{3}|1\rangle$ (occupied) and $|0\rangle$ (unoccupied) respectively. The final measurement layer projects the Pauli weight onto the basis $\{|0\rangle,\frac{1}{3}|1\rangle \}$ at each site. (b) Each transfer matrix (rectangular 4-leg tensor) describes a segment in the Markov process that converts the state of two adjacent sites into another in the 2-site Hilbert space, i.e., $\{|00\rangle,|01\rangle,|10\rangle,|11\rangle\}$. We denote an occupied site $|1\rangle$ as a blue right-mover (red left-mover) if its space-time coordinate follows $x+t=0(1)\bmod 2$, and unoccupied state $|0\rangle$ as lack of left or right movers.  The eight vertices with non-zero probability are shown. 
(c) demonstrates one configuration generated by Monte Carlo simulation according to the Markov process in (b), with only one site being occupied in the middle at $t=0$. Both left and right movers evolve at the speed of light and create a light-cone trajectory.}
\label{fig:circuit}
\end{figure}

    \section{Locally-scrambled Dual-unitary Dynamics}
To introduce the concept of a dual-unitary gate, we consider an arbitrary two-qubit gate $U$. In the computational basis, its space-time dual $\tilde{U}$ can be defined as 
\begin{equation}
\tilde{U}:=\sum_{ijkl}\ket{kl}\langle ik|U|jl\rangle\bra{ij},
\end{equation} 
The unitary $U$ is called dual-unitary when its space-time dual $\tilde{U}$ is also a unitary operator, satisfying $U^\dagger U=\tilde{U}^\dagger \tilde{U}=I$~\cite{bertiniExact2019}. For qubit systems, dual unitary gates are parameterized by four single-qubit unitaries $u_{\pm}$, $v_{\pm}$, and a single parameter $J$: 
\begin{myequation}
U(J)=(u_+\otimes u_-)e^{-i\frac{\pi}{4}(X_1X_2+Y_1Y_2)-iJZ_1Z_2}(v_+\otimes v_-).
\end{myequation}
Here, $\{X_j,Y_j,Z_j\}$ are the Pauli operators on the $j$-th qubit. Single-qubit gates provide local basis rotations but do not entangle different sites. The non-trivial entanglement arises from the XXZ coupling, where $J\in [0,\pi/4]$ characterizes the scrambling ability of the dual-unitary gate. For $J=\pi/4$, $U$ reduces to a SWAP operation, simply interchanging operators between two sites. Conversely, for $J=0$, it maximally spreading single-site operators to operators supported on both sites.

In this work, we study the operator dynamics of random dual-unitary circuits for qubit systems arranged in a brick-wall architecture, as shown in Fig.~\ref{fig:circuit} (a). The evolution operator is given by $U_{\text{tot}}(t_f)=\prod_{t=1}^{t_f} U_t $, where 
\begin{equation}\label{eqn:convention_evo}
U_t=
\begin{cases*}
\bigotimes_k U_{2k-1,2k} & if  $t \in \text{odd}$,  \\
\bigotimes_k U_{2k,2k+1} & if $t \in \text{even}$.
\end{cases*}
\end{equation}
Here, $U_{x,x'}$ represent two-qubit dual-unitary gates acting on nearest-neighbor qubits $x$ and $x'$. We note that throughout this paper, we use $U$ to denote the quantum evolution operator and $T$ the transfer matrix after mapping the dual-unitary quantum circuit to the classical Markov process. We sample each two-qubit gate $U$ independently in the locally-scrambled dual-unitary ensemble from the distribution $p(U|J)$ condition on the parameter $J$, and define the random unitary ensemble $\mathcal{E}_{J}$:
\begin{equation}
\mathcal{E}_{J}=\{U\sim p(U|J)=du_+du_-dv_+dv_-\},
\end{equation}
where $du_\pm$ and $dv_\pm$ represent the Haar measure of single-qubit rotations. This choice eliminates local basis dependence and is essential to mapping the full quantum dynamics to a classical Markov process~\cite{Kuo2020M1910.11351}, which then admits efficient Monte Carlo simulations, as we showed in this paper.

We are interested in the dynamical evolution of simple operators $O$. At time $t$, the Heisenberg evolution is given by $O(t+1)=U_{t}^\dagger O(t)U_{t}$. For later convenience, we represent the operator $O$ as a state $|O\rangle$ in a doubled Hilbert space, known as the Choi representation \cite{Jamiokowski1972L,Choi1975C}. On a complete set of orthonormal basis $\{\ket{i}\}$ of the Hilbert space, under the operator-state mapping, any generic operator $O=\sum_{ij}\ket{i}O_{ij}\bra{j}$ gets mapped to a corresponding doubled state as $\ket{O}={\color{black}2^{-L/2}}\sum_{ij}O_{ij}\ket{i}\otimes\ket{j}$. Here, $L$ denotes the system size.
In this language, the Heisenberg evolution reads $|O(t+1)\rangle=(U_t^\dagger \otimes U^T_t)|O(t)\rangle$. We choose a orthogonal basis in the operator space at site $x$ as $\{|I\rangle_x,|X\rangle_x,|Y\rangle_x, |Z\rangle_x \}$, where each basis state corresponds to having the operator $\{I,X,Y,Z\}$ acting on the qubit $x$, respectively. The operator wavefunction can be expanded as 
\begin{equation}
|O(t)\rangle=\sum_Pc_P(t) |P\rangle,
\end{equation}
where $|P\rangle=|P_{-L/2}P_{-L/2+1}...P_{L/2-1}\rangle $ with $P_x \in \{I,X,Y,Z\}$. We study the growth of a local operator that initially acts near the middle of the system (around $x \approx 0$), focusing on the time regime before it reaches the system boundaries under the operator dynamics, so that boundary effects remain irrelevant.

\subsection{Operator Density}
\label{sec:IIA}
We investigate the operator dynamics by monitoring both the evolution of operator density \cite{Nahum2018O1705.08975} and the operator entanglement entropy~\cite{bertiniOperatorI2020}. The operator density $\rho(x,t)$ is defined as the probability of finding a non-trivial operator on site $x$:
\begin{equation}
\rho(x,t)=\sum_{\{P|P_x\neq I\}}\overline{|c_P(t)|^2}\equiv \sum_{\{P|P_x\neq I\}}w_P(t).
\end{equation}
Here, the overline denotes the ensemble average over the realizations of random single-qubit gates in the dual-unitary ensemble. There are great simplications if we focus on locally-scrambled quantum circuits: the phase of $c_P(t)$ exhibits random fluctuations due to the random single-qubit rotations, and the evolution of $w_P(t)$ becomes Markovian~\cite{Kuo2020M1910.11351}. The evolution generated by each two-qubit gate $U_{xy}$ reads
\begin{equation}
w_{P}(t) ~\rightarrow~ w_P(t+1)={\sum_{P'}}^{\prime}T^{(2)}(P_xP_y,P_x'P_y')~w_{P'}(t),
 \end{equation}
where the summation $\sum'$ is restricted to the subspace where $P'_z=P_z$ for any $z\neq x,y$. The 16$\times$16 transfer matrix $T^{(2)}$ is computed in a two-qubit system by
\begin{equation}
\begin{aligned}
T^{(2)}(P,P')&:=\overline{|\langle P|(U^\dagger \otimes U^T)|P'\rangle|^2}\\&=\frac{1}{2^4}\int_{U\in\mathcal{E}_J} dU~\left|\text{tr}(PU^\dagger P'U)\right|^2.
\end{aligned}
\end{equation}
The completeness of Pauli operators guarantees the conservation of the probability, as $\sum_{P}T^{(2)}(P,P')=1$. 

Furthermore, we observe that the absence of local basis dependence suggests that $T^{(2)}(P,P')$ only depends on the support of $P$ and $P'$. As a consequence, $w_P(t)$ at $t>0$ also depends only on the support of $P$ rather than on the specific content of the Pauli string $P$. To eliminate the redundancy in $w_P$, we introduce a bit string $b=b_{-L/2}b_{-L/2+1}...b_{L/2-1}$ to denote the support of a Pauli string, where $b_x\in \{0,1\}$. We use $b_x=0$ to indicate the case where the site $x$ is not occupied by the operator and $b_x=1$ for the occupied case. Then, we can rewrite the Markov process in this occupation basis, where the transfer matrix can be reduced to a $4\times 4$ matrix $T$. Introduce $w_b(t)=3^{|b|}w_P(t)$, where $P$ is an arbitrary Pauli operator of support $b$, with $|b|$ denoting the number of occupied sites in the bit string $b$ (i.e.~the operator size), and the factor $3^{|b|}$ counts the total number of Pauli operators of the same support $b$. The operator density $\rho(x,t)$ can be expressed as
\begin{equation}
\rho(x,t)=\sum_{\{b|b_x=1\}}w_b(t).
\end{equation}
Defining $T(b,b')=3^{|b|}T^{(2)}(P,P')$, the evolution generated by each two-qubit gate remains
\begin{myequation}\label{eqn:wb_dynamics}
w_{b}(t) ~\rightarrow~ w_{b}(t+1)={\sum_{b'}}^{\prime}T(b_xb_y;b_x'b_y')~w_{b'}(t),
 \end{myequation}
 where the summation $\sum'$ is again restricted to the subspace where $b'_z=b_z$ for any $z\neq x,y$. The full Markovian dynamics is then represented by a brick-wall circuit of local transfer matrices $T$, as illustrated in Fig.~\ref{fig:circuit} (a). For the locally-scrambled dual-unitary ensemble, the explicit expression of $T(b,b')$ for each two-qubit gate has been computed~\cite{akhtar2024dual}, which reads
 \begin{myequation}
 \label{eqn:transfermatrix}
 T(b,b')=\begin{pmatrix}
 1&0&0&0\\
 0&0&1-\alpha&\alpha/3\\
 0&1-\alpha&0&\alpha/3\\
 0&\alpha&\alpha&1-2\alpha/3
 \end{pmatrix}_{bb'},
 \end{myequation}
 with parametrization $\alpha=\frac{2}{3}\cos^2(2J)\in[0,2/3]$ that controls the information scrambling power of the dual-unitary gate ensemble: $\alpha=0$ corresponds to the SWAP gate with no scrambling, and $\alpha=1$ corresponds to the fastest scrambling dual unitary gates. 

\subsection{Operator Entanglement}
\begin{figure}[t!]
\begin{center}
\includegraphics[scale=0.68]{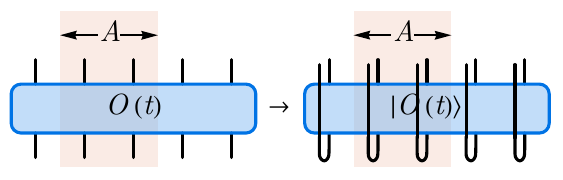}
\caption{Illustration of operator-state mapping and operator entanglement in region $A$.}
\label{fig:operator_entanglement}
\end{center}
\end{figure}

We also examine the growth of operator entanglement in the random dual-unitary circuit~\cite{bertiniOperatorI2020,bertiniOperatorII2020}. After mapping the operator to a state, the operator entanglement $O(t)$ for a subregion $A$ of the original qubit system is defined as the conventional entanglement entropy of the corresponding Choi state $|O(t)\rangle$ for the same subregion, as illustrated in Fig.~\ref{fig:operator_entanglement}.

More explicitly, we first construct the reduced density matrix in the operator space
\begin{equation}
\rho_A[O(t)]=\text{tr}_{\bar{A}} |O(t)\rangle \langle O(t)|.
\end{equation}
Here, $\bar{A}$ denotes the complement of the subregion $A$. We can compute either the von Neumann entropy $S_A(t)=-\text{tr}_A (\rho_A \log \rho_A)$ or the $n$-th R\'enyi entropy $S_A^{(n)}(t)=\frac{1}{1-n}\log \text{tr} (\rho_A^n)$. In this work, we focus on small subsystem size $|A|\ll |\bar{A}|$, where theoretical analysis becomes more tractable after the following approximations regarding the disorder average:
\begin{equation}\label{eqn:def_op_ent}
\begin{aligned}
\overline{S_A^{(n)}}(t)&=-\frac{1}{n-1}\overline{\log \text{tr} (\rho_A^n)}\approx -\frac{1}{n-1}\log \text{tr} (\overline{\rho_A^n})\\&\approx -\frac{1}{n-1}\log \text{tr} (\overline{\rho_A}^n).
\end{aligned}
\end{equation} 
In the first step, we approximate the disorder average of the entropy by the disorder average of $\overline{\text{tr} (\rho_A^n)}$. This approximation assumes that the corresponding variance is negligible within the ensemble of random circuits~\cite{hammaQuantum2012,Hayden2016H1601.01694}.  However, the calculation still requires the disorder average over 2$n$-replicas $U^{\otimes 2n}\otimes (U^\dagger)^{\otimes 2n}$, where the Schur-Weyl duality suggests that the effective degrees of freedom are elements of the $S_{2n}$ group. To further simplify the calculation, we assume that $\overline{S_A^{(n)}}$ is dominated by the ``replica diagonal'' contribution for small subsystems $|A|$, allowing us to replace $\overline{\rho_A^n}$ with $\overline{\rho_A}^n$, which constitutes our second step. It is known that these approximations become exact if we consider a system with qudits and take the limit of a large local Hilbert space dimension $(d\rightarrow \infty)$~\cite{bertiniOperatorI2020}. 

Under this approximation, we can express the second Rényi entropy in terms of $w_b(t)$.  As explained in Sec.~\ref{sec:IIA}, the disorder-averaged operator density matrix contains only the diagonal components:
$\bar{\rho}=\overline{|O(t)\rangle\langle O(t)|}=\sum_P w_P(t) |P\rangle\langle P|.$
Tracing out the subregion $\bar{A}$, we obtain the reduced density matrix
\begin{equation}\label{eqn:rhoA}
\begin{aligned}
 \overline{\rho_A}&=\sum_{P_A,P_{\bar{A}}} w_{P_AP_{\bar{A}}}(t) ~ |P_A\rangle \langle P_A|\\&\equiv \sum_{P_A} w_{P_A}(t) ~ |P_A\rangle \langle P_A|,
 \end{aligned}
\end{equation}
where we have decomposed $P=P_A\otimes P_{\bar{A}}$ and introduced the marginal distribution $w_{P_A}(t)=\sum_{P_{\bar{A}}}w_{P_AP_{\bar{A}}}(t)$. The purity of the system is then computed by
\begin{myequation}\label{eqn:SA2estimate}
\begin{aligned}
e^{-(n-1)\overline{S^{(n)}_A}(t)}\approx \sum_{P_A} w^n_{P_A}(t)= \sum_{b_A}\frac{1}{3^{|b_A|(n-1)}}w^n_{b_A}(t),
\end{aligned}
\end{myequation}
where $b_A$ denotes a bit string defined in region $A$.

\subsection{Monte Carlo Simulation}
Since $w_{b}(t)$ contains an exponentially large number of components, indexed by the bit string $b=\{b_x\}$ (with $b_x\in\{0,1\}$) that labels the operator support. Direct analytical or numerical study of the evolution of $w_{b}(t)$ following Eq.~\eqref{eqn:wb_dynamics} is impractical. To this end, we would like to use the Monte Carlo method to carry out important sampling and reduce the exponential computation to polynomial complexity. We may interpret the operator support $b$ as an occupation configuration of some fictitious ``particles'', and visualize the Markovian dynamics of operator evolution as the evolution of classical particles, where each lattice site $x$ is either empty ($b_x=0$) or occupied by a particle ($b_x=1$). The system is initialized according to the support of the initial operator $O$. The application of a transfer matrix in Eq.~\eqref{eqn:transfermatrix} on sites $x$ and $x+1$ is then implemented by the following local update rule, parametrized by the scrambling parameter $\alpha$ of the dual unitary gate:
\begin{enumerate}
\item If both sites are unoccupied, no update is needed.

\item If only site $x$ is occupied, we move the particle from site $x$ to site $x+1$. With a probability of $1-\alpha$, site $x$ is left empty. Otherwise, a new particle is added at site $x$.

\item If only site $x+1$ is occupied, we move the particle from site $x+1$ to site $x$. With a probability of $1-\alpha$, site $x+1$ is left empty. Otherwise, a new particle is added at site $x+1$.

\item If both sites are occupied, the configuration is kept unchanged with a probability $1-2\alpha/3$. Otherwise, one of the two particles is randomly removed, and the probability to remove either particle is $\alpha/3$.

\end{enumerate}
These update rules are graphically shown in Fig.~\ref{fig:circuit} (b) and they reveal a central feature of dual-unitary circuits with a brick-wall architecture. In the absence of scattering with other particles, a particle continues to move in a single spatial direction. Therefore, it is natural to distinguish lattice sites corresponding to right movers and left movers based on the parity of $x+t$. As an illustration, we depict right movers in blue and left movers in red in Fig.~\ref{fig:circuit} (c) for a typical evolution history. 

Starting from any initial configuration of the operator support (equivalently viewed as ``particle'' occupation configuration) at time $t=0$, we can stochastically generate the configuration $b$ at later time $t$ following the update rules described above. These update rules ensure that the underlying probability distribution of $b$ at time $t$ will precisely match $w_b(t)$, as if we can effectively sample $b$ from the distribution $w_b(t)$.

After collecting many independent samples, we can estimate the operator density $\rho(x,t)$ by counting the average particle number at the corresponding space-time position. On the other hand, the operator entanglement entropy $\overline{S_{A}^{(n)}}(t)$ is a non-linear function of the probability distribution $w_b(t)$. In the Monte Carlo simulation, we focus on the second R\'enyi entropy with R\'enyi index $n=2$. This first requires generating two independent configurations of classical particles at time $t$, labeled by their occupation bit strings $b^{(1)}$ and $b^{(2)}$. Then, we compute 
\begin{equation}
I(b^{(1)},b^{(2)})=\prod_{x\in A}3^{-b_x^{(1)}}\delta_{b^{(1)}_xb^{(2)}_x}
\label{EEestimator}
\end{equation}
as an estimator of Eq.~\eqref{eqn:SA2estimate}. Here, $\delta_{b^{(1)}_xb^{(2)}_x}$ is the Kronecker delta symbol that enforces $b^{(1)}$ and $b^{(2)}$ have identical configurations inside the entanglement region $A$ of interest, otherwise the quantity $I(b^{(1)},b^{(2)})$ will be zero. Averaging $I(b^{(1)},b^{(2)})$ over many pairs of samples gives $\av{e^{-\overline{S_A^{(2)}}}}$, and then $\overline{S_A^{(2)}}=-\ln \av{I(b^{(1)},b^{(2)})}$. 

As one can see from Eq.~\eqref{EEestimator}, a non-zero $I$ is sampled only if $b_x^{(1)}=b_x^{(2)}, \forall x\in A$, which is an exponentially rare event to happen. This makes the sampling exponentially inefficient for estimating the operator entanglement. In practice, we mitigate this issue by computing the probability distribution at time $t$ condition on the configuration generated at time $t-1$, where $t$ denotes the time slice where operator properties are measured. Recall that each local transfer matrix $T$ acts on one pair of left and right movers, and is unrelated to the others pair of movers in the same time slice due to the brick-wall architecture. Therefore, for every two sites (a pair of movers), we can explicitly compute the probability of all four possible states in the 2-site state space occurring at $t$, conditioned on the current state at $t-1$. Let's denote the particle configuration at time $t$ by the bit string $b=\{b_x\}$, and that at time $t-1$ by the bit string $b'=\{b'_x\}$. Then we can compute the quantity $I$ as a function of $b'^{(1)}$ and $b'^{(2)}$, which are independently sampled in two replica up to the $t-1$ time step, 
\begin{equation}
I(b^{\prime(1)}, b^{\prime(2)})=\prod_{2k\in A}\sum_{b_{2k-1},b_{2k}}3^{-(b_{2k-1}+b_{2k})}\tilde{w}_{b|b^{\prime(1)}}\tilde{w}_{b|b^{\prime(2)}}
\end{equation}
which amounts to directly summing over all the final-time configurations $b$ based on the conditional probability $\tilde{w}_{b|b^{\prime(1)}}\tilde{w}_{b|b^{\prime(2)}}$. Here $\tilde{w}_{b|b'}$ corresponds to the conditional probability to generate $b$ given $b'$ in a single replica, which is given by (the tensor product) of the transfer matrix $T(b,b')$, defined in Eq.~\eqref{eqn:transfermatrix},
\begin{equation}
\tilde{w}_{b|b'}=\prod_{k}T(b_{2k-1}b_{2k};b'_{2k-1}b'_{2k-1}),
\end{equation}
assuming the last time step is an odd step without lost of generality. In this way, averaging $I(b'^{(1)},b'^{(2)})$ also gives unbiased estimation of the operator entanglement $\overline{S_A^{(2)}}=-\ln \langle I(b'^{(1)},b'^{(2)})\rangle$. Since the summation of configurations $b$ can be factorized to that on each pair of sites at $(2k-1,2k)$, which can computed efficiently, we can now obtain a finite real number $I(b'^{(1)},b'^{(2)})$ from each Monte Carlo sampling instead of waiting an exponentially long time for  $\delta_{b^{(1)}_xb^{(2)}_x}=1$ to occur. Similar approach in spirit, to overcome the exponential observable with log-normal distribution to a typical observable with normal distribution~\cite{liaoControllable2023,zhangIntegral2024,zhouIncremental2024}, has been employed in the recent computation of entanglement entropy and other exponential observables in the quantum many-body systems within quantum Monte Carlo and Tensor Network setup ~\cite{hastingsMeasuring2010,improvingLuitz2014,song2024resummation,demidioEntanglement2020,zhaoScaling2022,zhaoMeasuring2022,song2024Extracting,wangAnalog2024,zhangIntegral2024,songEvolution2025,wangEntanglement2025}. {All code used for generating the results in this work can be accessed online~\cite{github}.
}

\section{Results}


In the following, we will investigate the operator dynamics and operator entanglement growth in dual-unitary circuits using the Monte Carlo sampling approach described above, and compare the results with mean-field theoretical understandings.

For the operator dynamics, we focusing on two initial operators: \( O = Z_0 Z_1 \) and \( O = Z_1 \). For \( O = Z_0 Z_1 \), we find that the operator density evolution closely follows the mean-field approximation, with our analytical solutions in both the long-time limit and near the light-cone matching well with Monte Carlo simulations. Conversely, for \( O = Z_1 \), the operator density near the left light-cone deviates significantly from mean-field predictions due to strong correlations arising from the stochastic emission time of the first left-moving operator. By incorporating a conditional probability framework, we derive analytical expressions that accurately capture these correlations, achieving excellent agreement with numerical results. 

Additionally, we explore operator entanglement growth in three different subregions near the light-cone and identify entanglement transitions as the parameter \( \alpha \) varies. Our analytical predictions of critical values for these transitions are validated by Monte Carlo simulations, which has not been observed in previous works~\cite{bertiniExact2019,bertiniOperatorI2020,bertiniOperatorII2020,akhtar2024dual}. Our discovery therefore highlights the sensitivity of entanglement growth to initial conditions and system parameters. 

Overall, our findings demonstrate the intricate interplay between operator dynamics, correlations, and entanglement scaling in dual-unitary circuits, enhancing the understanding of information spreading, thermalization, and quantum information scrambling in quantum many-body systems.

\subsection{Operator Density Evolution}
\label{sec:IIIA}

\begin{figure}[t!]
\includegraphics[width=\columnwidth]{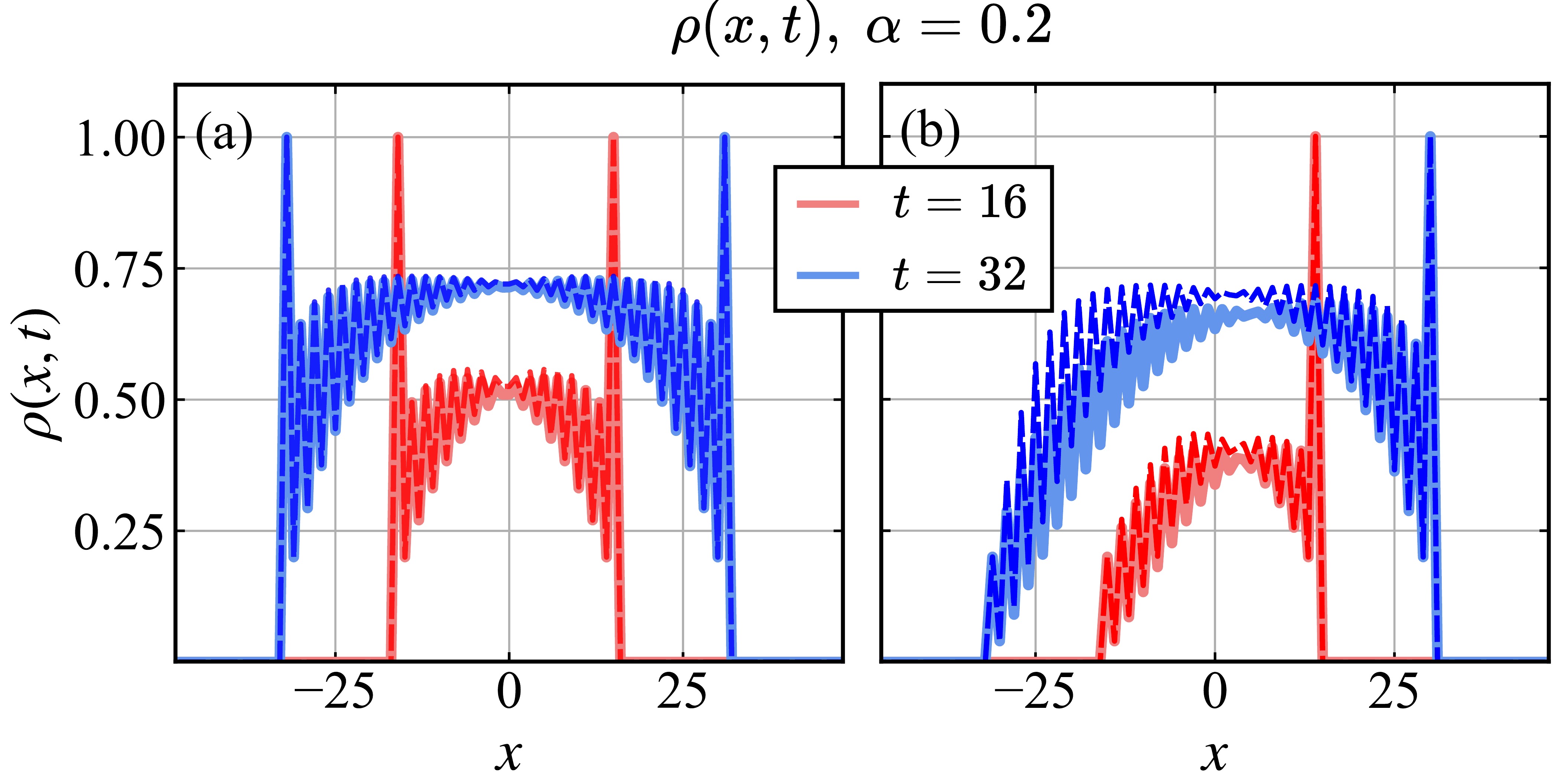}
\caption{\textbf{Comparing density profile obtained from the recurrence relation in Eq.~\eqref{eq:recur} with Monte Carlo data at $\alpha=0.2$.} Initially, panel (a) has two sites occupied in the middle and panel (b) has a single site occupied in the middle. In both panels, solid lines represent the Monte Carlo data with $\sim 10^7$ samplings such that the error bar is negligible in the figure. The dashed lines are generated from the recurrence relation Eq.~\eqref{eq:recur}. The MF prediction matches well with the MC data with the two-site initial condition at any time $t$, while it deviates from the MC result at large $t$ with a single-site initial condition.}
\label{fig:compare}
\end{figure}

We begin with the analysis of operator density evolution. For illustration, we focus on two different initial operators: $O=Z_0 Z_1 $ (two-site operator) or $O=Z_1$ (single-site operator), which turns out to show different features. According to our convention in Eq.~\eqref{eqn:convention_evo}, $Z_0$ represents a left mover, while $Z_1$ corresponds to a right mover. We present typical results of the operator density evolution in Fig.~\ref{fig:compare} (a) and (b) using solid lines for $\alpha=0.2$ and $t=16,32$. The data presented is averaged over $10^7$ Monte Carlo steps such that error bars are negligible from bare-eye observation.

\subsubsection{Balanced operator growth from two-site operator}

Let us first focus on the scenario with $O=Z_0 Z_1 $. On both the left ($x+t=0$) and right ($x-t=1$) light-cones, the operator density is 1, consistent with the ballistic transport of the initial operator. Within this light-cone, the operator density continues to increase due to the creation of new particles at finite splitting probability $\alpha$. In the long-time limit, the operator density on each site $x$ approaches $3/4$, indicating that each single-qubit Pauli operator $\{I,X,Y,Z\}$ appears with equal probability, as expected for generic chaotic quantum systems~\cite{Ippoliti2023O2212.11963}. To gain a more quantitative understanding, we adopt the zeroth-order mean-field approximation~\cite{akhtar2024dual}, which approximates the joint distribution $w_b$ of the operator support $b$ as product of independent Bernoulli distribution of $b_x$ on each site $x$, parametrized by the operator density $\rho(x,t)$ as
\begin{equation}
\label{eqn:approximation}
w_b(t)\approx \prod_x \left[b_x\rho(x,t)+(1-b_x)(1-\rho(x,t))\right].
\end{equation}
This mean-field approximation neglects any correlation between different sites. The evolution of $\rho(x,t)$ is then derived by applying a layer of transfer matrix $T$ and computing the post-evolution density $\rho(x,t+1)$~\cite{akhtar2024dual}. For a transfer matrix $T$ applied to sites $x$ and $x+1$, the evolution reads
\begin{myequation}
\begin{aligned}
&\rho(x,t+1)=\rho(x+1,t)+\alpha\rho(x,t)\Big(1-\frac{4}{3}\rho(x+1,t)\Big),\\
&\rho(x+1,t+1)=\rho(x,t)+\alpha\rho(x+1,t)\Big(1-\frac{4}{3}\rho(x,t)\Big).
\label{eq:recur}
\end{aligned}
\end{myequation}
The evolution generated by these equations is plotted in Fig.~\ref{fig:compare} (a) using dashed lines. The results match the Monte Carlo simulation with high accuracy for $O=Z_0 Z_1 $. Additionally, we provide analytical solutions in two limits:
\begin{enumerate}
\item In the long-time limit $t\gg 1$ for a fixed lattice site $x$, we can neglect the spatial dependence of the operator density and expand $\rho(t)=3/4-\delta \rho(t)$. The evolution then becomes 
\begin{equation}
\delta \rho(t+1)=(1-\alpha)\delta\rho(t)+O(\delta \rho^2). 
\end{equation}
This equation can be solved exactly, which gives 
\begin{equation}\label{eqn:relaxation}
\rho(t)\approx \frac{3}{4}-Ae^{-\lambda t},\ \ \ \text{with } \lambda=-\ln(1-\alpha).
\end{equation}

\item Now, we consider the near-light-cone steady state for the operator density. Specifically, we examine the case where $d=x+t\sim O(1)$ or $d=t-x+1\sim O(1)$ with $t\gg 1$. Numerical results suggest the existence of a steady-state distribution $\rho^{\text{LC}}_d$, where LC stands for ``light-cone'', which satisfies $\rho^{\text{LC}}_{-1}=0$, $\rho^{\text{LC}}_0=1$ and
\begin{equation}
\begin{aligned}
&\rho^{\text{LC}}_{2k}=\rho^{\text{LC}}_{2k}+\alpha \rho^{\text{LC}}_{2k-1}\Big(1-\frac{4}{3}\rho^{\text{LC}}_{2k}\Big),\\
&\rho^{\text{LC}}_{2k+1}=\rho^{\text{LC}}_{2k-1}+\alpha \rho^{\text{LC}}_{2k}\Big(1-\frac{4}{3}\rho^{\text{LC}}_{2k-1}\Big).
\end{aligned}
\end{equation}
Using the first equation, we find 
\begin{equation}\label{eqn:2site_LC_2k}
\rho_{2k}^{\text{LC}}=3/4
\end{equation}
for any $k>0$. Substituting this into the second equation, we have ($k\geq 0$)
\begin{equation}\label{eqn:2site_LC}
\rho^{\text{LC}}_{2k+1}=\frac{3}{4}-\left(\frac{3}{4}-\alpha\right)(1-\alpha)^k.
\end{equation}
\end{enumerate}

\begin{figure}[t!]
\includegraphics[width=0.9\columnwidth]{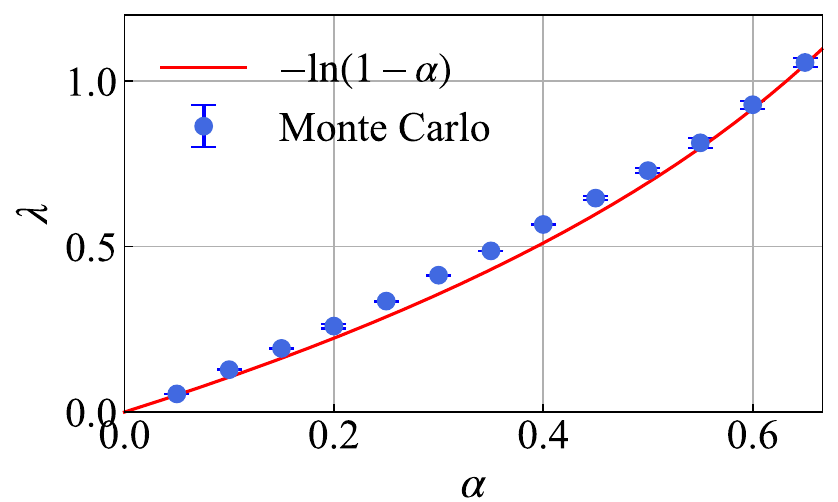}
\caption{\textbf{$\lambda$ in Eq.~\eqref{eqn:relaxation} as a function of $\alpha$.} The blue dots are fitted from the MC data of $\rho(0,t)$ with $O=Z_0Z_1$ initial condition. The red solid line denotes the mean-field solution in the long-time limit $t\gg 1$.}
\label{fig:lambda}
\end{figure}

\begin{figure}[t!]
\includegraphics[width=0.95\columnwidth]{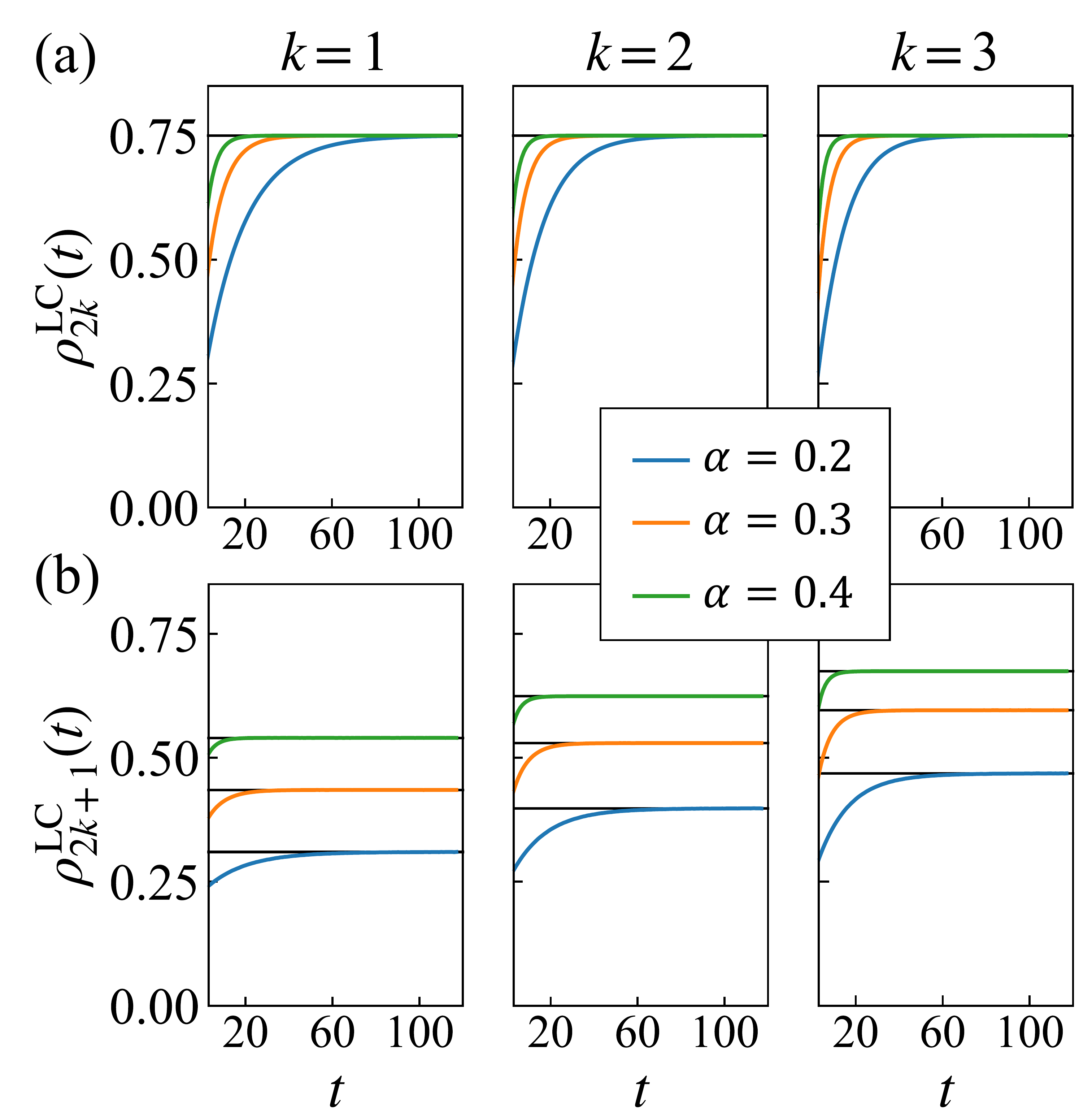}
\caption{\textbf{Converge of the near-light-cone operator density $\rho^{\mathrm{LC}}$ obtained from Monte Carlo sampling with $O=Z_0Z_1$ initial condition.} Panels (a) and (b) display the density evolution of left movers ($\rho_{2k}^{\mathrm{LC}}$) and right movers ($\rho_{2k+1}^{\mathrm{LC}}$) respectively. Colored lines represent the Monte Carlo data with $\sim 10^7$ samplings, whereas the horizontal black lines in panels (a) and (b) denote the mean-field result $\rho_{2k}^{\mathrm{LC}}=3/4$ and Eq.~\eqref{eqn:2site_LC} at corresponding $k$ values, respectively.}
\label{fig:2site_rho_LC}
\end{figure}

We test both predictions through the Monte Carlo sampling of the $\rho(x,t)$ at various $\alpha$ for $O=Z_0Z_1$. Since $\rho$ becomes spatially independent at $t\gg 1$, we focus on the density evolution at $x=0$ without loss of generality. By fitting the density evolution with Eq.~\eqref{eqn:relaxation}, we find an exponential growth coefficient $\lambda$ that is consistent with the analytical solution in Eq.~\eqref{eqn:relaxation}, as shown in Fig.~\ref{fig:lambda}. For the near-light-cone steady state, MC data of both $\rho_{2k}^{\mathrm{LC}}$ and $\rho_{2k+1}^{\mathrm{LC}}$ exponentially converges to the MF prediction in Eq.~\eqref{eqn:2site_LC} at large-enough $t$, as exemplified in Fig.~\ref{fig:2site_rho_LC} (a) and (b) for $k=1,2,3$ at $\alpha=0.2,0.3,0.4$.

\subsubsection{Skewed operator growth from single-site operator}

We then consider the operator density evolution for $O=Z_1$, as shown in Fig.~\ref{fig:compare} (b). In this case, the right light-cone ($x-t=1$) has an operator density of 1, while the density on the left light-cone $(x+t=2)$ is $\alpha$, determined by the probability of operator emission at the first step. Unlike the previous example with $O=Z_0Z_1$, numerical results show that the operator density evolution near the left light-cone differs significantly from a direct mean-field solution. This discrepancy arises because the mean-field analysis is valid only when the operator occupations on different sites are nearly independent, allowing correlations to be neglected. However, strong correlations exist near the left light-cone for $O=Z_1$, and this can be explained in the following way.

We consider the conditional probability $\rho(x,t|t_0)$ of the full operator dynamics, where $t_0=1,2,...$ labels the time when the initial operator emits a left mover for the first time.
This left mover becomes the leftmost particle and will never disappear in subsequent evolution. The probability distribution of $t_0$ is given by
\begin{equation}\label{eqn:t0prior}
p(t_0)=\alpha (1-\alpha)^{t_0-1}.
\end{equation} 
After the emission of the first left mover, we have two nearest neighbor particles, and the evolution of operator density matches previous discussions for $O=Z_0Z_1$. This results in
\begin{equation}
\rho(x,t~|t_0)=\rho(x-t_0,t-t_0)_{ZZ}.
\end{equation}
Here, $\rho(x,t)_{ZZ}$ is the operator density for $O=Z_0Z_1$. Specifically, we have $\rho(x,t|t_0)=0$ if $x+t< 2t_0$ and $\rho(x,t|t_0)=1$ if $x+t=2t_0$, indicating a long-range correlation between different sites near the left light-cone after summing up contributions from different $t_0$. Therefore, a naive mean-field theory does not apply. However, since it has been established that $\rho(x,t)_{ZZ}$ can be described by the mean-field equation, we can still make predictions using a superposition of mean-field results:
\begin{equation}
\rho(x,t)=\sum_{t_0=1}^\infty p(t_0) \rho(x-t_0,t-t_0)_{ZZ}.
\end{equation}
Since $p(t_0)$ exponentially localizes at small $t_0$, evident from Eq.~\eqref{eqn:t0prior}, it does not change the long-time relaxation to $\rho(x)=3/4$ and the steady-state operator density near the right light-cone, which is still given by Eq.~\eqref{eqn:2site_LC}. For the steady-state near the left light-cone, the dependence on $t_0$ becomes significant. Defining $\tilde{d}=x+t-2\sim O(1)$, we have $\tilde{\rho}^{\text{LC}}_{0}=\alpha$,
\begin{myequation}
\begin{aligned}
\label{eqn:1site_LC_2k}
\tilde{\rho}^{\text{LC}}_{2k}&=\frac{3}{4}\sum_{t_0=1}^{k}p(t_0)+p(k+1)\\
&=\frac{3}{4}\left(1-(1-\alpha)^{k-1}\right)+\alpha (1-\alpha)^k.
\end{aligned}
\end{myequation}
and 
\begin{myequation}
\begin{aligned}
\label{eqn:1site_LC_2k+1}
\tilde{\rho}^{\text{LC}}_{2k+1}&=\sum_{t_0=1}^{k+1}p(t_0)\left(\frac{3}{4}-\left(\frac{3}{4}-\alpha\right)(1-\alpha)^{k+1-t_0}\right)\\
&=\frac{3}{4}-\frac{3}{4}(1-\alpha)^k\left[1-\alpha\left(\frac{4}{3}\alpha+k\left(-1+\frac{4}{3}\alpha\right)\right)\right].
\end{aligned}
\end{myequation}
for any $k>0$. 

\begin{figure}[t!]
\includegraphics[width=0.95\columnwidth]{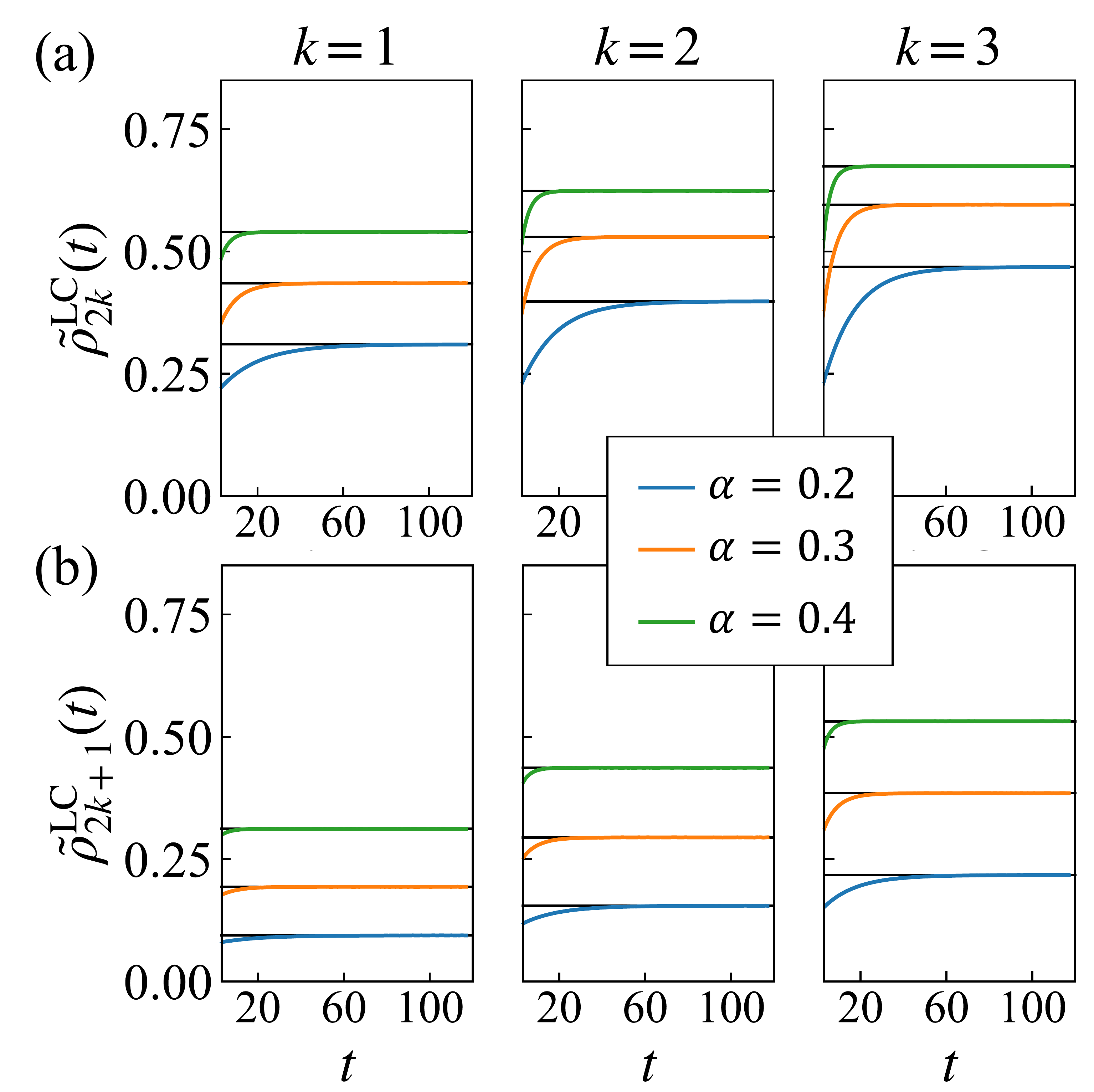}
\caption{\textbf{Converge of the near-left-light-cone operator density $\tilde{\rho}^{\mathrm{LC}}$ obtained from Monte Carlo sampling with $O=Z_1$ initial condition.} Panels (a) and (b) display the density evolution of left movers ($\tilde{\rho}_{2k}^{\mathrm{LC}}$) and right movers ($\tilde{\rho}_{2k+1}^{\mathrm{LC}}$) respectively. Colored lines represent the Monte Carlo data with $\sim 10^7$ samplings, whereas the horizontal black lines in panels (a) and (b) denote the mean-field results in Eqs.~\eqref{eqn:1site_LC_2k} and~\eqref{eqn:1site_LC_2k+1} at corresponding $k$ values, respectively.}
\label{fig:1site_rho_LC}
\end{figure}

Again, we test the above prediction near the left light-cone with $O=Z_1$ initial condition by directly comparing with Monte Carlo simulations, as exemplified in Fig.~\ref{fig:1site_rho_LC} (a) and (b) for $k=1,2,3$ at $\alpha=0.2,0.3,0.4$. The operator density near the left light-cone for both left and right movers converges to the mean-field results in Eqs.~\eqref{eqn:1site_LC_2k} and~\eqref{eqn:1site_LC_2k+1} as $t$ increases. 

\subsubsection{Summary of results on operator density}

\begin{figure}[t!]
\begin{center}
\includegraphics[scale=0.75]{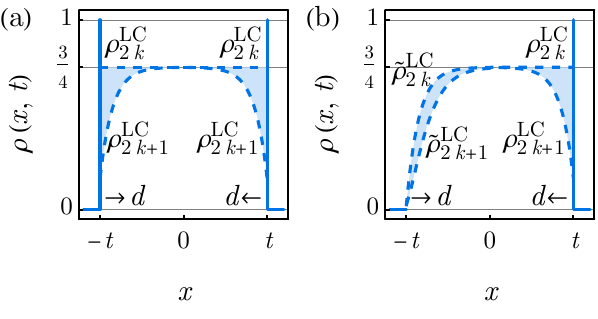}
\begin{tabular}{lll}
\hline
Bound & Expression & Verification \\
\hline
$\rho^\mathrm{LC}_{2k}$ & Eq.~\eqref{eqn:2site_LC_2k} & Fig.~\ref{fig:2site_rho_LC}(a) \\ 
$\rho^\mathrm{LC}_{2k+1}$ & Eq.~\eqref{eqn:2site_LC} & Fig.~\ref{fig:2site_rho_LC}(b) \\
$\tilde{\rho}^\mathrm{LC}_{2k}$ & Eq.~\eqref{eqn:1site_LC_2k} & Fig.~\ref{fig:1site_rho_LC}(a) \\
$\tilde{\rho}^\mathrm{LC}_{2k+1}$ & Eq.~\eqref{eqn:1site_LC_2k+1} & Fig.~\ref{fig:1site_rho_LC}(b) \\
\hline
\end{tabular}
\caption{Schematic plot the operator density distribution $\rho(x,t)$ in the long-time limit, for operator dynamics starting with (a) $O=Z_0$ and (b) $O=Z_0Z_1$. The shaded area indicates the operator density oscillates between the upper and lower bounds alternatively. The bounds near the light-cone edge are denoted as $\rho^\mathrm{LC}_{d}$ or $\tilde{\rho}^\mathrm{LC}_{d}$, with $d$ being the distance away from the light-cone edge. Their mean-field expressions and verifications with Monte Carlo simulations are summarized in the following table.}
\label{fig:density}
\end{center}
\end{figure}

In this subsection, we investigated the operator growth dynamics in dual-unitary circuits with a brick-wall arrangement of two-qubit gates, characterized by the gate parameter \( \alpha \). Near the center of the light cone of the growing operator, the operator density converges exponentially to \( \frac{3}{4} \), following \( \rho(t) = \frac{3}{4}-A e^{-\lambda t} \), with the rate \( \lambda = -\ln(1-\alpha) \) determined by \( \alpha \). In the long-time limit, the operator density will converge to the profile depicted in Fig.~\ref{fig:density}, which oscillates between upper and lower bounds. We derived analytical expressions for these bounds based on a mean-field-like theory, where operators spread like diffusive ``particles'', and correlations among these particles are retained only to the minimal necessary order. Their expressions and verifications are summarized in the table within Fig.~\ref{fig:density}.

\subsection{Operator Entanglement Growth}
\label{sec:IIIB}
Next, we study the dynamics of operator entanglement, focusing on the steady-state entanglement for subregions near the light-cone starting from the single-site operator $O=Z_1$. As established in Sec.~\ref{sec:IIIA}, the results obtained near the right light-cone for $O=Z_1$ should also apply to both light-cone edges for the two-site operator $O=Z_0Z_1$, so we will not further discuss the operator entanglement dynamics for $O=Z_0Z_1$, and we will focus our discussion on the case of $O=Z_1$ only. Given the initial operator $O=Z_1$, as it evolves under the dual unitary circuit, the operator spreads into a light-cone in the spacetime. 

\begin{figure}[t!]
\begin{center}
\includegraphics[scale=0.75]{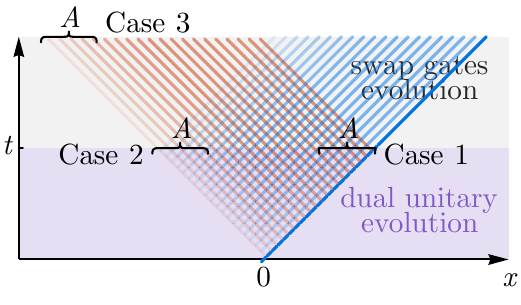}
\caption{Illustration of three different choices of the operator entanglement region $A$ in the spacetime. Assuming a single-site operator at the spacetime origin evolve under dual unitary circuits for sufficiently long time, Cases 1 and 2 concerns the operator entanglement near the two inequivalent light-cone edges. Case 3 consider the operator entanglement of the left movers only, which can be effectively separated from right movers by a subsequent circuit of swap gates.}
\label{fig:region choices}
\end{center}
\end{figure}

We consider three different cases with different choices for subregion $A$ to study the operator entanglement entropy, after evolving $O=Z_1$ by the dual unitary circuit for time $t$, as shown in Fig.~\ref{fig:region choices}:
\begin{enumerate}
\item $A$ contains consecutive sites from the right light-cone edge.

\item $A$ contains consecutive sites from the left light-cone edge.

\item $A$ contains consecutive left movers from the left light-cone edge. 
\end{enumerate} 
It worth mention that the Case 3 can be viewed as the operator entanglement on the left light-cone after further evolving the operator by a brick-wall circuit of swap gates (i.e.~quenching dual unitary gates to swap gates after time $t$), such that the left- and right-movers are separated. We will be most interested in the scaling of operator entanglement entropy with the region size $l_A$ in the long-time limit $t\to\infty$, as the operator has been sufficiently scrambled by the dual unitary circuit to exhibit equilibrium behavior.

\subsubsection{Case 1: from the right light-cone edge}
We begin with the Case 1, which allows for a direct mean-field description. With the mean-field approximation Eq.~\eqref{eqn:approximation}, we can estimate the $n$th R\'enyi operator entanglement Eq.~\eqref{eqn:SA2estimate} as 
\begin{myequation}
\overline{S^{(n)}_{A,1}}(t)\approx -\frac{1}{n-1}\sum_{x\in A}\ln\left((1-\rho(x,t))^n+\rho^n(x,t)/3^{n-1}\right). 
\end{myequation}

\begin{figure}[t!]
\includegraphics[width=0.9\columnwidth]{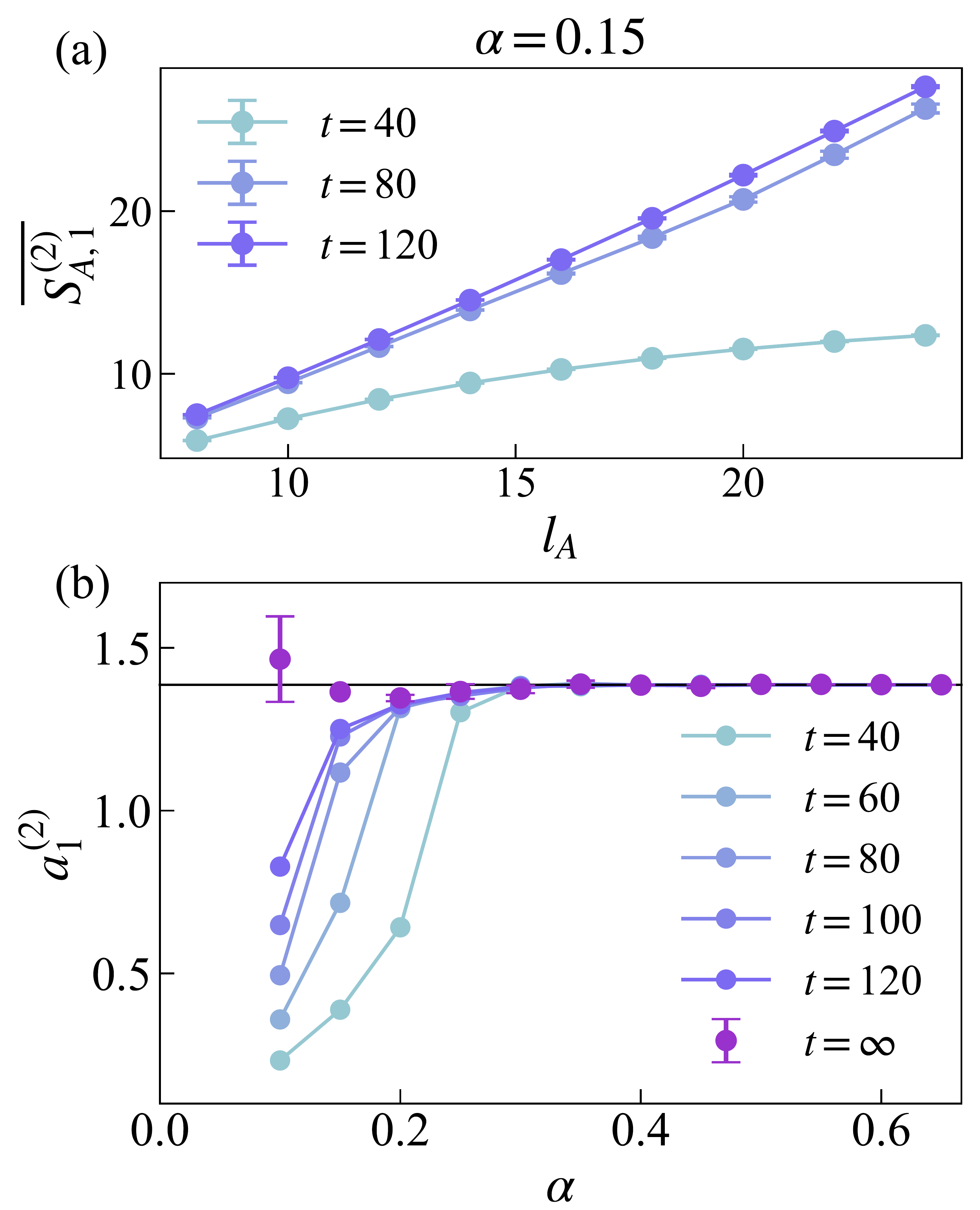}
\caption{\textbf{Scaling the operator entanglement entropy for Case 1.} (a) The linear growth of the operator entanglement entropy with respect to the entanglement region length $l_A$ at $\alpha=0.15$ as an example. Notice that the evolution time at small $\alpha$ should be long enough to manifest the linear growth. (b) The fitted leading volume law coefficient as a function of $\alpha$ obtained at various evolution times $t$. $\aone$ values at $t=\infty$ are extrapolated from the finite-$t$ results at each $\alpha$ with a power-law fitting. The horizontal black line represents the maximal growth speed $2\ln2$. The entanglement entropy data is averaged over $\sim 10^8$ samplings.}
\label{fig:head_ee}
\end{figure}

Near the right light-cone, the steady-state operator density, given by Eq.~\eqref{eqn:2site_LC}, approaches $\rho=3/4$ within a length scale of $-2/\log(1-\alpha)$. Therefore, to extract the volume law coefficient of operator entanglement, defined as $$a^{(n)}\equiv \lim_{t\gg l_A \gg 1}\overline{S^{(n)}_A}/l_A,$$ we can safely approximate $\rho^{\text{LC}}_d=3/4$. This leads to the expression $\overline{S^{(n)}_{A,1}}\approx (2\ln 2)~l_A+O(1)$, yielding 
\begin{equation}\label{eqn:right_EE}
a_1^{(n)}=2\ln 2.
\end{equation}
Since the local operator space has a dimension of 4, this result demonstrates that operator entanglement near the right light-cone is nearly maximal. We demonstrate the maximum growth of the operator entanglement entropy at $n=2$, i.e., second R\'enyi entropy in Fig.~\ref{fig:head_ee}. As one can see in Fig.~\ref{fig:head_ee} (a), linear scaling of $\renyione$ against the entanglement region $l_A$ (volume law) is observed only after a long enough evolution time $t$, especially at small $\alpha$ values. The fitted volume law coefficients $\aone$ against $\alpha$ at different $t$ are presented in Fig.~\ref{fig:head_ee} (b). At large $\alpha$, $\aone$ is found to be maximum $2\ln 2$, even at small $t$. However, $\aone$ drifts drastically against $t$ at small $\alpha$. Therefore, we extrapolate the $t\to \infty$ limit from finite-$t$ $\aone$ values, denoted as the purple dots with error bars in Fig.~\ref{fig:head_ee} (b). The extrapolated values demonstrate a maximum growth of the operator entanglement entropy for all $\alpha \in [0,2/3]$ at $t\to \infty$, consistent with the mean-field analysis above.

\subsubsection{Case 2: from the left light-cone edge}

\begin{figure}[t!]
\includegraphics[width=0.9\columnwidth]{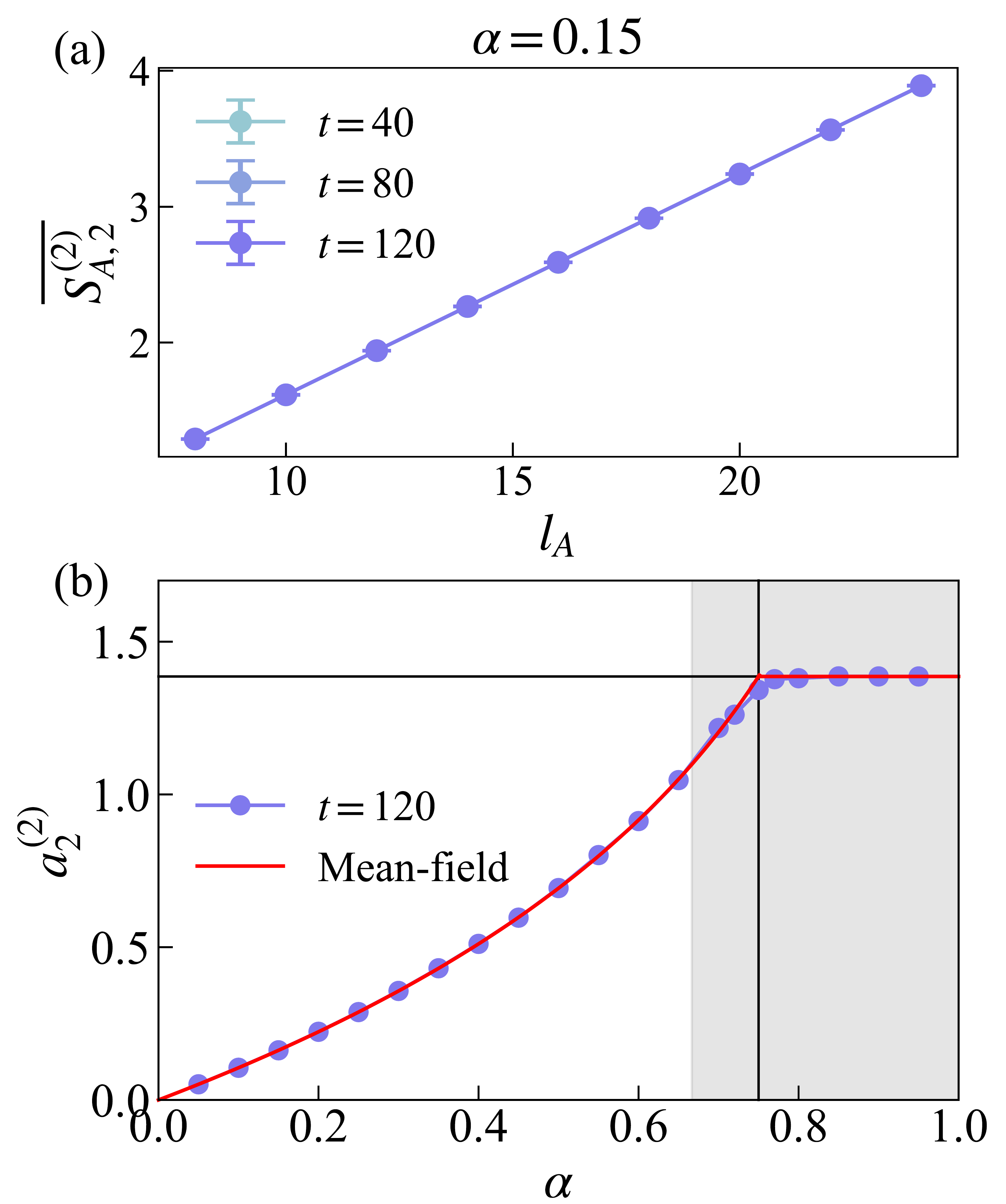}
\caption{\textbf{Scaling the operator entanglement entropy for Case 2.} (a) The fitted linear growth of the operator entanglement entropy with respect to the entanglement region length $l_A$ at $\alpha=0.15$ as an example. The operator entanglement entropy measured at different times is nearly identical such that the markers overlap with each other. (b) The leading volume law coefficient as a function of $\alpha$ obtained at various evolution times $t=120$. A transition between a maximal and a sub-maximal growth of the entanglement entropy manifests. The horizontal black line represents the maximal growth speed $2\ln2$, whereas the vertical one indicates the transition point $\alpha=3/4$. The shaded area marks the unphysical region where $\alpha$ exceeds $2/3$. The entanglement entropy data computed by averaging over $\sim 10^8$ samples.}
\label{fig:tail_ee}
\end{figure}

Now, let us consider Case 2 near the left light-cone. Similar to the calculation of the operator density, we introduce the conditional joint distribution function $w_{P_A}(t|t_0)$ given an emission of a left-mover for the first time at $t_0$ and express the reduced density matrix as follows:
\begin{myequation}
\overline{\rho_A}=\sum_{t_0}p(t_0)\sum_{P_{A}^{(t_0)}}w_{P_A^{(t_0)}}(t|t_0)~|P_A^{(t_0)}\rangle \langle P_A^{(t_0)}|.
\end{myequation}
Here, $w_{P_A}(t|t_0)$ matches the joint distribution with the initial operator $O=Z_0Z_1$, with shifted space-time indices. An important observation is that $t_0$ determines the position of the leftmost particle. Therefore, for $t_0,t_0'\leq (l_A+1)/2$ and $t_0\neq t_0'$, we have $\langle P_A^{(t_0)} | P_A^{(t_0')}\rangle =0$. On the other hand, for $t_0>(l_A+1)/2$, the subsystem $A$ contains only the identity operator $|I\rangle$. Consequently, we can rewrite the reduced density matrix as 
\begin{myequation}
\begin{aligned}
\overline{\rho_A}=&\sum_{t_0\leq \frac{l_A+1}{2}}p(t_0)\sum_{P_{A}^{(t_0)}}w_{P_A^{(t_0)}}(t|t_0)~|P_A^{(t_0)}\rangle \langle P_A^{(t_0)}|\\&+\sum_{t_0>\frac{l_A+1}{2}}p(t_0) |I\rangle.
\end{aligned}
\end{myequation}
Here, all different terms are orthogonal. Taking the mean-field approximation for the conditional distribution, we can express the operator entanglement entropy as 
\begin{myequation}\label{eqn:Sn}
\begin{aligned}
e^{-(n-1)\overline{S^{(n)}_{A,2}}}=&\sum_{1\leq t_0\leq \frac{l_A+1}{2}}p^n(t_0) e^{-(n-1)\overline{S^{(n)}_{\tilde{A},1}}}\\&+\Bigg(\sum_{t_0>\frac{l_A+1}{2}}p(t_0)\Bigg)^n.
\end{aligned}
\end{myequation}
In this equation, we have identified the contribution from $\sum_{P_{A}^{(t_0)}}w_{P_A^{(t_0)}}(t|t_0)~|P_A^{(t_0)}\rangle \langle P_A^{(t_0)}|$ as the operator entanglement of a subsystem $\tilde{A}$, which contains $l_A-2t_0+2$ sites near the left light-cone for the initial operator $O=Z_0Z_1$. The corresponding result matches the findings of Case 1 for $O=Z_1$. 

To extract the volume-law coefficient, we use $\overline{S^{(n)}_{\tilde{A},1}}\approx (l_A-2t_0+2)\times 2\ln2$. Moreover, for large $l_A$, we can neglect the last term in Eq. \eqref{eqn:Sn}. The result reads
\begin{myequation}
e^{-(n-1)\overline{S^{(n)}_{A,2}}}=\alpha^n\sum_{1\leq t_0\leq \frac{l_A+1}{2}}(1-\alpha)^{n(t_0-1)}4^{-(n-1)(l_A-2t_0+2)}.
\end{myequation}
Instead of a direct calculation of the summation, we can estimate by extracting the $t_0$ dependence, which scales as $\beta^{t_0}$ with $\beta:=2^{4(n-1)}(1-\alpha)^n$. When we have $\beta>1$, the dominant contribution is from $t_0 \approx (l_A+1)/2 $, while for $\beta<1$, the dominant contribution comes from $t_0 \approx 0$. Therefore, we have
\begin{myequation}
\label{eqn:left_EE}
a_2^{(n)}=
\begin{cases*}
-\frac{n}{2(n-1)}\ln(1-\alpha) & if  $\alpha<1-2^{-\frac{4(n-1)}{n}}$,  \\
2\ln 2& if $\alpha>1-2^{-\frac{4(n-1)}{n}}$.
\end{cases*}
\end{myequation}
This suggests a possible transition in operator entanglement as we tune $\alpha$. Unfortunately, since the mapping from the original quantum circuits to the Markovian process requires $\alpha\leq 2/3$, the transition point $\alpha_c=1-2^{-\frac{4(n-1)}{n}}$ is never accessible for integers $n \geq 2$. On the other hand, taking the limit $n=1$ leads to $\alpha_c=0$. A non-trivial entanglement transition at finite $\alpha$ only exist for $1<n<(1-(\log_2 3)/4)^{-1}\approx 1.66$.

From Eq.~\eqref{eqn:left_EE}, the transition for the second R\'enyi entanglement, $n=2$ appears at $\alpha_c=3/4$. Although $\alpha=3/4$ is out of the legitimate domain of the dual unitary gate parameter, we can nonetheless perform Monte Carlo sampling in the entire range $\alpha \in [0,1]$. Fig.~\ref{fig:tail_ee} (a) demonstrates the volume law growth of the entanglement entropy. In this case, entanglement entropy converges very quickly against the evolution time $t$ such that the data measured at $t=40$ is almost identical to that measured at $t=120$, in contrast to the strong time-dependent feature observed in \textit{Case 1}. A transition at $\alpha_c=3/4$ is successfully detected by the volume law coefficient $\atwo$: below $\alpha=3/4$, $\atwo$ increases as $-\ln (1-\alpha)$ consistent with mean-field analysis and remains the maximum value when $\alpha>3/4$. 

\subsubsection{Case 3: from the left light-cone edge, focusing on left-movers only}

\begin{figure}[t!]
\includegraphics[width=0.9\columnwidth]{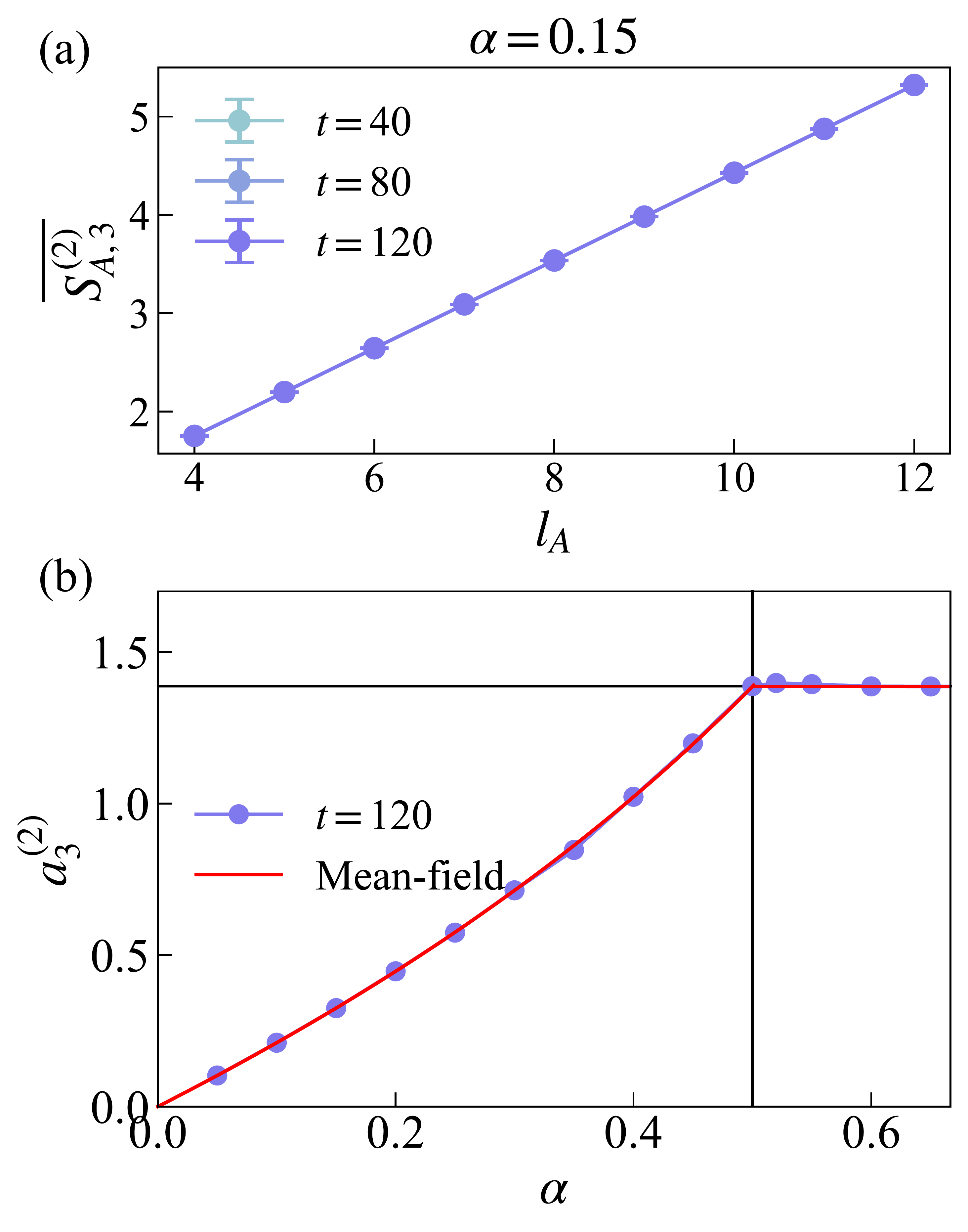}
\caption{\textbf{Scaling the operator entanglement entropy for Case 3.} (a) The linear growth of the operator entanglement entropy with respect to the entanglement region containing only left-movers at $\alpha=0.15$ as an example. The operator entanglement entropy measured at different times is nearly identical such that the markers overlap with each other. (b) The fitted leading volume law coefficient as a function of $\alpha$ obtained at various evolution times $t=120$. A transition between a maximal and a sub-maximal growth of the entanglement entropy manifests. The horizontal black line represents the maximal growth speed $2\ln2$, whereas the vertical one indicates the transition point $\alpha=1/2$. The entanglement entropy data is averaged over $\sim 10^8$ samplings.}
\label{fig:tail_left_ee}
\end{figure}

We can construct a scenario where an entanglement transition can be observed at a finite $\alpha$ within the physical range of the gate parameter. This is described in Case 3, where we consider only the subsystem $A$ of left movers. Following the analysis of Case 2, the only difference is a reduction of $\overline{S^{(n)}_{\tilde{A},1}}$ by a factor of two. Therefore, we expect
\begin{myequation}
e^{-(n-1)\overline{S^{(n)}_{A,3}}}\approx \alpha^n\sum_{1\leq t_0\leq l_A+1}(1-\alpha)^{n(t_0-1)}4^{-(n-1)(l_A-t_0+1)}.
\end{myequation}
This results in a change in the critical $\alpha$:
\begin{equation}\label{eqn:left_EE_L}
a_3^{(n)}=
\begin{cases*}
-\frac{n}{n-1}\ln(1-\alpha) & if  $\alpha<1-2^{-\frac{2(n-1)}{n}}$,  \\
2\ln 2& if $\alpha>1-2^{-\frac{2(n-1)}{n}}$.
\end{cases*}
\end{equation}
An entanglement transition exists for $1<n<(1-(\log_2 3)/2)^{-1}\approx 4.82$. In particular, we find $\alpha_c=1/2$ for the second R\'enyi operator entanglement, which is also verified via numerical simulation in Fig.~\ref{fig:tail_left_ee} (b). The growing behavior of $\athree$ at the sub-maximal growth region, i.e., $\alpha<1/2$, also agrees with the mean-field prediction, which reads $-2\ln (1-\alpha)$ at $n=2$.

\subsubsection{Summary of results on operator entanglement}

\begin{figure}[t!]
\begin{center}
\includegraphics[scale=0.7]{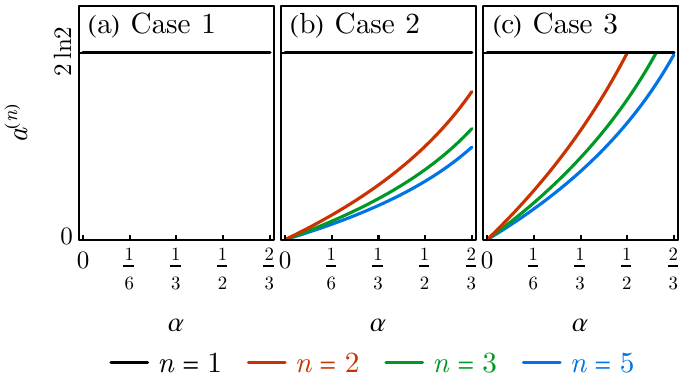}
\begin{tabular}{cll}
\hline
Coefficient & Expression & Verification ($n=2$) \\
\hline
$a_1^{(n)}$ & Eq.~\eqref{eqn:right_EE} & Fig.~\ref{fig:head_ee}(b) \\ 
$a_2^{(n)}$ & Eq.~\eqref{eqn:left_EE} & Fig.~\ref{fig:tail_ee}(b) \\
$a_3^{(n)}$ & Eq.~\eqref{eqn:left_EE_L} & Fig.~\ref{fig:tail_left_ee}(b) \\
\hline
\end{tabular}
\caption{Summary of the volume-law coefficient $a^{(n)}$ of the $n$th-R\'enyi operator entanglement entropy, depending on the dual unitary gate parameter $\alpha$ for three cases of entanglement regions: (a) from the right light-cone edge, (b) from the left light-cone edge, (c) from the left light-cone edge focusing on left-movers only. Maximal to sub-maximal volume-law transition in operator entanglement only happens in the last case.}
\label{fig:volume laws}
\end{center}
\end{figure}

In this subsection, we studied the dynamics of operator entanglement in dual-unitary circuits, examining three distinct cases based on different subregions near the light-cone with the initial operator \( O = Z_1 \), as illustrated in Fig.~\ref{fig:region choices}. Across all three cases, the $n$th-R\'enyi operator entanglement entropy $\overline{S_{A}^{(n)}}(t)$, as defined in Eq.~\eqref{eqn:def_op_ent}, exhibits a volume-law scaling on both edges of the light cone in the long-time limit $t\to\infty$, i.e.
\begin{myequation}
\overline{S_{A}^{(n)}}\simeq a^{(n)}l_A+\cdots,
\end{myequation}
growing linearly with the subregion size \( l_A \). However, the dependence of the volume-law coefficient $a^{(n)}$ on the gate parameter \( \alpha \) and the R\'enyi index $n$ differs between the cases, as summarized in Fig.~\ref{fig:volume laws}.

The transition from maximal to sub-maximal volume-law growth in operator entanglement is observed specifically in Case 3, when focusing on the left movers in the left light-cone entanglement region. This result highlights how the gate parameter \( \alpha \) influence the degree of operator entanglement among the leading left movers emitted from the expanding operator. Such operator entanglement transitions, characterized by the abrupt drop in the volume-law coefficient at a critical value of \( \alpha \), have been previously discussed in the literature~\cite{bertiniExact2019,bertiniOperatorI2020,bertiniOperatorII2020}. Notably, Ref.~\cite{bertiniOperatorI2020} first identified the operator entanglement transition in a setup akin to our Case 2 and derived the following formula for qubit systems:
\begin{myequation}
a^{(n)}=
\begin{cases*}
-\frac{n}{n-1}\ln|\lambda|^2 & if  $1>|\lambda|^2>2^{-\frac{2(n-1)}{n}}$,  \\
2\ln 2& otherwise,
\end{cases*}
\end{myequation}
where \( \lambda \) is the leading non-trivial eigenvalue of a transfer matrix defined therein. However, the explicit dependence of \( \lambda \) on the gate parameter was not fully established. Our study builds upon this foundation by demonstrating that this behavior actually aligns precisely with our Case 3 scenario when we identify \( |\lambda|^2 = 1 - \alpha = 1 - \frac{2}{3} \cos^2(2J) \). 

As a result, we clarify the appropriate entanglement region setup necessary to observe the operator entanglement transition in dual-unitary circuits, providing a mean-field formula for the volume-law coefficient directly in terms of the gate parameter. This result is supported by Monte Carlo simulations, which show excellent agreement with our theoretical predictions. 

\section{Discussion}
\label{sec:discussion}
In this manuscript, we systematically investigate the operator density evolution and entanglement entropy scaling in dual-unitary circuits by mapping the dynamics to a classical Markov process, enabling efficient Monte Carlo simulations. For operator density, we demonstrate that near the center of the light cone, the operator density converges exponentially to a steady value of \( \frac{3}{4} \), with a rate governed by the gate parameter \( \alpha \). In the long-time limit, we derive analytical expressions for the operator density profile's upper and lower bounds, which are verified through simulations. 

For operator entanglement, we analyze three distinct subregions near the light cone and find that the operator entanglement entropy consistently follows a volume-law scaling. However, the volume-law coefficient's dependence on \( \alpha \) differs across cases. Notably, we identify a transition from maximal to sub-maximal volume-law growth when focusing on left movers near the left light cone. This transition, characterized by a sudden drop in the volume-law coefficient at a critical \( \alpha \), aligns with prior theoretical predictions in the literature~\cite{bertiniExact2019,bertiniOperatorI2020,bertiniOperatorII2020}. Additionally, we extend this understanding by providing a direct mapping of the volume-law coefficient to the gate parameter, supported by mean-field theory and validated through Monte Carlo simulations. Our work not only clarifies the conditions under which operator entanglement transitions occur but also offers a comprehensive framework for studying operator dynamics and entanglement in dual-unitary circuits.

Our result could have implications in \emph{classical shadow tomography}~\cite{Huang2020P2002.08953} and can be extended to random quantum circuits with symmetries. Classical shadow tomography is a recently developed method for efficiently predicting many properties of quantum systems using a limited number of measurements~\cite{Ohliger2013E1204.5735,Guta2020F1809.11162,Huang2020P2002.08953,Elben2020M2007.06305}. By understanding operator dynamics and entanglement growth, we can improve the efficiency of classical shadow protocols, leading to more accurate reconstructions of quantum states and observables in many-body systems \cite{Huang2021P2011.01938,Huang2021I2101.02464,Huang2021P2106.12627,Huang2021E2103.07510,Levy2021C2110.02965,Acharya2021S2105.05992,Hu2022H2102.10132,Bu2022C2202.03272,Seif2022S2203.07309,Hu2022L2203.07263,Chau-Nguyen2022O2205.08990,Hadfield2022M2006.15788,Huang2022Q2112.00778,Enshan-Koh2022C2011.11580,Hao-Low2022C2208.08964,Bertoni2022S2209.12924,Arienzo2022C2211.09835,Hu2023C2107.04817,Akhtar2023S2209.02093,zhangHolographic2024,Akhtar2024M2308.01653,Hu2024D2402.17911,Liu2023P2311.00695,Shen2024E2409.13691,Ippoliti2024L,Ippoliti2024C2305.10723,Chen2024O2404.19105}. More specifically, the advantage of using dual unitary circuit in classical shadow tomography has been recently discussed in Ref.~\cite{akhtar2024dual}.


An interesting future direction is to extend our stochastic simulation framework to random quantum circuits with symmetries, which play a crucial role in modeling realistic quantum systems that conserve quantities such as particle number or spin~\cite{Khemani2018O1710.09835,Rakovszky2018D1710.09827,Hunter-Jones2018O1812.08219,agrawalEntanglement2022,folignoNonequilibrium2024,Chang2024D2408.15325,Fisher2023R2207.14280}. Incorporating symmetries into operator dynamics affects the spreading and entanglement properties, offering deeper insights into the role of conserved quantities in quantum information scrambling and thermalization. Our approach provides a pathway to study these effects efficiently, bridging the gap between solvable models and more complex, realistic systems.

\begin{acknowledgments}
We thank Yimu Bao, Tian-Gang Zhou for discussions. MHS, TTW and ZYM acknowledge the support from the Research Grants Council (RGC) of Hong Kong (Project Nos. 17301721, AoE/P-701/20, 17309822, HKU C7037-22GF, 17302223, 17301924), the ANR/RGC Joint Research Scheme sponsored by RGC of Hong Kong and French National Research Agency (Project No. A\_HKU703/22), the GD-NSF (No. 2022A1515011007). PZ is supported by the Shanghai Rising-Star Program under grant number 24QA2700300, the NSFC under grant 12374477, and the Innovation Program for Quantum Science and Technology 2024ZD0300101. YZY is supported by a startup fund from UCSD. We thank HPC2021 system under the Information Technology Services, The University of Hong Kong, as well as the Beijing PARATERA Tech CO.,Ltd. (URL: https://cloud.paratera.com) for providing HPC resources that have contributed to the research results reported within this paper. ZYZ, YZY and ZYM acknowledge the hospitality of Kavli Institute for Theoretical Physics (KITP) and thank the organizers of the KITP program ``Correlated Gapless Quantum Matter''  where this work was initiated. This research was supported in part by grant NSF PHY-2309135 to the KITP.
\end{acknowledgments}

\bibliographystyle{unsrtnat}
\bibliography{main.bbl}

\begin{thebibliography}{95}
\providecommand{\natexlab}[1]{#1}
\providecommand{\url}[1]{\texttt{#1}}
\expandafter\ifx\csname urlstyle\endcsname\relax
  \providecommand{\doi}[1]{doi: #1}\else
  \providecommand{\doi}{doi: \begingroup \urlstyle{rm}\Url}\fi

\bibitem[{Ho} and {Abanin}(2017)]{Ho2017E1508.03784}
Wen~Wei {Ho} and Dmitry~A. {Abanin}.
\newblock {Entanglement dynamics in quantum many-body systems}.
\newblock \emph{Physical Review B}, 95\penalty0 (9):\penalty0 094302, March
  2017.
\newblock \doi{10.1103/PhysRevB.95.094302}.

\bibitem[{Nahum} et~al.(2017){Nahum}, {Ruhman}, {Vijay}, and
  {Haah}]{Nahum2017Q1608.06950}
Adam {Nahum}, Jonathan {Ruhman}, Sagar {Vijay}, and Jeongwan {Haah}.
\newblock {Quantum Entanglement Growth under Random Unitary Dynamics}.
\newblock \emph{Physical Review X}, 7\penalty0 (3):\penalty0 031016, July 2017.
\newblock \doi{10.1103/PhysRevX.7.031016}.

\bibitem[{Nahum} et~al.(2018){Nahum}, {Vijay}, and
  {Haah}]{Nahum2018O1705.08975}
Adam {Nahum}, Sagar {Vijay}, and Jeongwan {Haah}.
\newblock {Operator Spreading in Random Unitary Circuits}.
\newblock \emph{Physical Review X}, 8\penalty0 (2):\penalty0 021014, April
  2018.
\newblock \doi{10.1103/PhysRevX.8.021014}.

\bibitem[{von Keyserlingk} et~al.(2018){von Keyserlingk}, {Rakovszky},
  {Pollmann}, and {Sondhi}]{von-Keyserlingk2018O1705.08910}
C.~W. {von Keyserlingk}, Tibor {Rakovszky}, Frank {Pollmann}, and S.~L.
  {Sondhi}.
\newblock {Operator Hydrodynamics, OTOCs, and Entanglement Growth in Systems
  without Conservation Laws}.
\newblock \emph{Physical Review X}, 8\penalty0 (2):\penalty0 021013, April
  2018.
\newblock \doi{10.1103/PhysRevX.8.021013}.

\bibitem[{Khemani} et~al.(2018){Khemani}, {Vishwanath}, and
  {Huse}]{Khemani2018O1710.09835}
Vedika {Khemani}, Ashvin {Vishwanath}, and David~A. {Huse}.
\newblock {Operator Spreading and the Emergence of Dissipative Hydrodynamics
  under Unitary Evolution with Conservation Laws}.
\newblock \emph{Physical Review X}, 8\penalty0 (3):\penalty0 031057, July 2018.
\newblock \doi{10.1103/PhysRevX.8.031057}.

\bibitem[{Chan} et~al.(2018){Chan}, {De Luca}, and
  {Chalker}]{Chan2018S1712.06836}
Amos {Chan}, Andrea {De Luca}, and J.~T. {Chalker}.
\newblock {Solution of a Minimal Model for Many-Body Quantum Chaos}.
\newblock \emph{Physical Review X}, 8\penalty0 (4):\penalty0 041019, October
  2018.
\newblock \doi{10.1103/PhysRevX.8.041019}.

\bibitem[{Zhou} and {Chen}(2019)]{Zhou2019O1805.09307}
Tianci {Zhou} and Xiao {Chen}.
\newblock {Operator dynamics in a Brownian quantum circuit}.
\newblock \emph{Physical Review E}, 99\penalty0 (5):\penalty0 052212, May 2019.
\newblock \doi{10.1103/PhysRevE.99.052212}.

\bibitem[{Qi} et~al.(2019){Qi}, {Davis}, {Periwal}, and
  {Schleier-Smith}]{Qi2019M1906.00524}
Xiao-Liang {Qi}, Emily~J. {Davis}, Avikar {Periwal}, and Monika
  {Schleier-Smith}.
\newblock {Measuring operator size growth in quantum quench experiments}.
\newblock \emph{arXiv e-prints}, art. arXiv:1906.00524, June 2019.
\newblock \doi{10.48550/arXiv.1906.00524}.

\bibitem[{von Keyserlingk} et~al.(2022){von Keyserlingk}, {Pollmann}, and
  {Rakovszky}]{von-Keyserlingk2022O2111.09904}
Curt {von Keyserlingk}, Frank {Pollmann}, and Tibor {Rakovszky}.
\newblock {Operator backflow and the classical simulation of quantum
  transport}.
\newblock \emph{Physical Review B}, 105\penalty0 (24):\penalty0 245101, June
  2022.
\newblock \doi{10.1103/PhysRevB.105.245101}.

\bibitem[{Schuster} and {Yao}(2022)]{Schuster2022O2208.12272}
Thomas {Schuster} and Norman~Y. {Yao}.
\newblock {Operator Growth in Open Quantum Systems}.
\newblock \emph{arXiv e-prints}, art. arXiv:2208.12272, August 2022.
\newblock \doi{10.48550/arXiv.2208.12272}.

\bibitem[{Bohrdt} et~al.(2017){Bohrdt}, {Mendl}, {Endres}, and
  {Knap}]{Bohrdt2017S1612.02434}
A.~{Bohrdt}, C.~B. {Mendl}, M.~{Endres}, and M.~{Knap}.
\newblock {Scrambling and thermalization in a diffusive quantum many-body
  system}.
\newblock \emph{New Journal of Physics}, 19\penalty0 (6):\penalty0 063001, June
  2017.
\newblock \doi{10.1088/1367-2630/aa719b}.

\bibitem[{Kukuljan} et~al.(2017){Kukuljan}, {Grozdanov}, and
  {Prosen}]{Kukuljan2017W1701.09147}
Ivan {Kukuljan}, Sa{\v{s}}o {Grozdanov}, and Toma{\v{z}} {Prosen}.
\newblock {Weak quantum chaos}.
\newblock \emph{Physical Review B}, 96\penalty0 (6):\penalty0 060301, August
  2017.
\newblock \doi{10.1103/PhysRevB.96.060301}.

\bibitem[{Xu} and {Swingle}(2019)]{Xu2019L1805.05376}
Shenglong {Xu} and Brian {Swingle}.
\newblock {Locality, Quantum Fluctuations, and Scrambling}.
\newblock \emph{Physical Review X}, 9\penalty0 (3):\penalty0 031048, July 2019.
\newblock \doi{10.1103/PhysRevX.9.031048}.

\bibitem[{Parker} et~al.(2019){Parker}, {Cao}, {Avdoshkin}, {Scaffidi}, and
  {Altman}]{Parker2019A1812.08657}
Daniel~E. {Parker}, Xiangyu {Cao}, Alexander {Avdoshkin}, Thomas {Scaffidi},
  and Ehud {Altman}.
\newblock {A Universal Operator Growth Hypothesis}.
\newblock \emph{Physical Review X}, 9\penalty0 (4):\penalty0 041017, October
  2019.
\newblock \doi{10.1103/PhysRevX.9.041017}.

\bibitem[{Kuo} et~al.(2020){Kuo}, {Akhtar}, {Arovas}, and
  {You}]{Kuo2020M1910.11351}
Wei-Ting {Kuo}, A.~A. {Akhtar}, Daniel~P. {Arovas}, and Yi-Zhuang {You}.
\newblock {Markovian entanglement dynamics under locally scrambled quantum
  evolution}.
\newblock \emph{Physical Review B}, 101\penalty0 (22):\penalty0 224202, June
  2020.
\newblock \doi{10.1103/PhysRevB.101.224202}.

\bibitem[{Akhtar} and {You}(2020)]{Akhtar2020M2006.08797}
A.~A. {Akhtar} and Yi-Zhuang {You}.
\newblock {Multiregion entanglement in locally scrambled quantum dynamics}.
\newblock \emph{Physical Review B}, 102\penalty0 (13):\penalty0 134203, October
  2020.
\newblock \doi{10.1103/PhysRevB.102.134203}.

\bibitem[{Zhang} and {Zhai}(2023)]{Zhang2023U2305.02356}
Ren {Zhang} and Hui {Zhai}.
\newblock {Universal Hypothesis of Autocorrelation Function from Krylov
  Complexity}.
\newblock \emph{arXiv e-prints}, art. arXiv:2305.02356, May 2023.
\newblock \doi{10.48550/arXiv.2305.02356}.

\bibitem[{Liu} et~al.(2023{\natexlab{a}}){Liu}, {Tang}, and
  {Zhai}]{Liu2023K2207.13603}
Chang {Liu}, Haifeng {Tang}, and Hui {Zhai}.
\newblock {Krylov complexity in open quantum systems}.
\newblock \emph{Physical Review Research}, 5\penalty0 (3):\penalty0 033085,
  August 2023{\natexlab{a}}.
\newblock \doi{10.1103/PhysRevResearch.5.033085}.

\bibitem[{Bu{\v{c}}a}(2023)]{Buca2023U2301.07091}
Berislav {Bu{\v{c}}a}.
\newblock {Unified Theory of Local Quantum Many-Body Dynamics: Eigenoperator
  Thermalization Theorems}.
\newblock \emph{Physical Review X}, 13\penalty0 (3):\penalty0 031013, July
  2023.
\newblock \doi{10.1103/PhysRevX.13.031013}.

\bibitem[Zhang and Gu(2023)]{Zhang:2022fma}
Pengfei Zhang and Yingfei Gu.
\newblock {Operator size distribution in large N quantum mechanics of Majorana
  fermions}.
\newblock \emph{JHEP}, 10:\penalty0 018, 2023.
\newblock \doi{10.1007/JHEP10(2023)018}.

\bibitem[Zhang and Yu(2023)]{Zhang:2022atf}
Pengfei Zhang and Zhenhua Yu.
\newblock {Dynamical Transition of Operator Size Growth in Quantum Systems
  Embedded in an Environment}.
\newblock \emph{Phys. Rev. Lett.}, 130\penalty0 (25):\penalty0 250401, 2023.
\newblock \doi{10.1103/PhysRevLett.130.250401}.

\bibitem[{Gottesman}(1998)]{Gottesman1998Tquant-ph/9807006}
Daniel {Gottesman}.
\newblock {The Heisenberg Representation of Quantum Computers}.
\newblock \emph{arXiv e-prints}, art. quant-ph/9807006, July 1998.
\newblock \doi{10.48550/arXiv.quant-ph/9807006}.

\bibitem[{Vidal}(2004)]{Vidal2004Equant-ph/0310089}
Guifr{\'e} {Vidal}.
\newblock {Efficient Simulation of One-Dimensional Quantum Many-Body Systems}.
\newblock \emph{Physical Review Letters}, 93\penalty0 (4):\penalty0 040502,
  July 2004.
\newblock \doi{10.1103/PhysRevLett.93.040502}.

\bibitem[Haegeman et~al.(2011)Haegeman, Cirac, Osborne,
  Pi\ifmmode~\check{z}\else \v{z}\fi{}orn, Verschelde, and
  Verstraete]{haegemanTime2011}
Jutho Haegeman, J.~Ignacio Cirac, Tobias~J. Osborne, Iztok
  Pi\ifmmode~\check{z}\else \v{z}\fi{}orn, Henri Verschelde, and Frank
  Verstraete.
\newblock Time-dependent variational principle for quantum lattices.
\newblock \emph{Phys. Rev. Lett.}, 107:\penalty0 070601, Aug 2011.
\newblock \doi{10.1103/PhysRevLett.107.070601}.

\bibitem[Haegeman et~al.(2016)Haegeman, Lubich, Oseledets, Vandereycken, and
  Verstraete]{haegemanUnifying2016}
Jutho Haegeman, Christian Lubich, Ivan Oseledets, Bart Vandereycken, and Frank
  Verstraete.
\newblock Unifying time evolution and optimization with matrix product states.
\newblock \emph{Phys. Rev. B}, 94:\penalty0 165116, Oct 2016.
\newblock \doi{10.1103/PhysRevB.94.165116}.

\bibitem[Muth et~al.(2011)Muth, Unanyan, and Fleischhauer]{muthDynamical2011}
Dominik Muth, Razmik~G. Unanyan, and Michael Fleischhauer.
\newblock Dynamical simulation of integrable and nonintegrable models in the
  heisenberg picture.
\newblock \emph{Phys. Rev. Lett.}, 106:\penalty0 077202, Feb 2011.
\newblock \doi{10.1103/PhysRevLett.106.077202}.

\bibitem[Alba et~al.(2019)Alba, Dubail, and Medenjak]{albaOperator2019}
V.~Alba, J.~Dubail, and M.~Medenjak.
\newblock Operator entanglement in interacting integrable quantum systems: The
  case of the rule 54 chain.
\newblock \emph{Phys. Rev. Lett.}, 122:\penalty0 250603, Jun 2019.
\newblock \doi{10.1103/PhysRevLett.122.250603}.

\bibitem[{Jonay} et~al.(2018){Jonay}, {Huse}, and {Nahum}]{jonayCoarse2018}
Cheryne {Jonay}, David~A. {Huse}, and Adam {Nahum}.
\newblock {Coarse-grained dynamics of operator and state entanglement}.
\newblock \emph{arXiv e-prints}, art. arXiv:1803.00089, February 2018.
\newblock \doi{10.48550/arXiv.1803.00089}.

\bibitem[{Zhou} and {Nahum}(2019)]{Zhou2019E1804.09737}
Tianci {Zhou} and Adam {Nahum}.
\newblock {Emergent statistical mechanics of entanglement in random unitary
  circuits}.
\newblock \emph{Physical Review B}, 99\penalty0 (17):\penalty0 174205, May
  2019.
\newblock \doi{10.1103/PhysRevB.99.174205}.

\bibitem[Chen and
  Kudler-Flam(2025)]{chen2024freeindependencenoncrossingpartition}
Hyaline~Junhe Chen and Jonah Kudler-Flam.
\newblock {Free independence and the noncrossing partition lattice in
  dual-unitary quantum circuits}.
\newblock \emph{Phys. Rev. B}, 111\penalty0 (1):\penalty0 014311, 2025.
\newblock \doi{10.1103/PhysRevB.111.014311}.

\bibitem[{de Groot} et~al.(2022){de Groot}, {Turzillo}, and
  {Schuch}]{de-Groot2022S2112.04483}
Caroline {de Groot}, Alex {Turzillo}, and Norbert {Schuch}.
\newblock {Symmetry Protected Topological Order in Open Quantum Systems}.
\newblock \emph{Quantum}, 6:\penalty0 856, November 2022.
\newblock \doi{10.22331/q-2022-11-10-856}.

\bibitem[{Ma} and {Wang}(2023)]{Ma2023A2209.02723}
Ruochen {Ma} and Chong {Wang}.
\newblock {Average Symmetry-Protected Topological Phases}.
\newblock \emph{Physical Review X}, 13\penalty0 (3):\penalty0 031016, July
  2023.
\newblock \doi{10.1103/PhysRevX.13.031016}.

\bibitem[Nahum et~al.(2022)Nahum, Roy, Vijay, and Zhou]{Nahum2022R2205.11544}
Adam Nahum, Sthitadhi Roy, Sagar Vijay, and Tianci Zhou.
\newblock Real-time correlators in chaotic quantum many-body systems.
\newblock \emph{Phys. Rev. B}, 106:\penalty0 224310, Dec 2022.
\newblock \doi{10.1103/PhysRevB.106.224310}.

\bibitem[{Nie} et~al.(2019){Nie}, {Nozaki}, {Ryu}, and {Tian
  Tan}]{Nie2019S1812.00013}
Laimei {Nie}, Masahiro {Nozaki}, Shinsei {Ryu}, and Mao {Tian Tan}.
\newblock {Signature of quantum chaos in operator entanglement in 2d CFTs}.
\newblock \emph{Journal of Statistical Mechanics: Theory and Experiment},
  9\penalty0 (9):\penalty0 093107, September 2019.
\newblock \doi{10.1088/1742-5468/ab3a29}.

\bibitem[{Kudler-Flam} et~al.(2020){Kudler-Flam}, {Nozaki}, {Ryu}, and
  {Tan}]{Kudler-Flam2020Q1906.07639}
Jonah {Kudler-Flam}, Masahiro {Nozaki}, Shinsei {Ryu}, and Mao~Tian {Tan}.
\newblock {Quantum vs. classical information: operator negativity as a probe of
  scrambling}.
\newblock \emph{Journal of High Energy Physics}, 2020\penalty0 (1):\penalty0
  31, January 2020.
\newblock \doi{10.1007/JHEP01(2020)031}.

\bibitem[Bertini et~al.(2020{\natexlab{a}})Bertini, Kos, and
  Prosen]{bertiniOperatorII2020}
Bruno Bertini, Pavel Kos, and Tomaz Prosen.
\newblock {Operator Entanglement in Local Quantum Circuits II: Solitons in
  Chains of Qubits}.
\newblock \emph{SciPost Phys.}, 8:\penalty0 068, 2020{\natexlab{a}}.
\newblock \doi{10.21468/SciPostPhys.8.4.068}.

\bibitem[Bertini et~al.(2020{\natexlab{b}})Bertini, Kos, and
  Prosen]{bertiniOperatorI2020}
Bruno Bertini, Pavel Kos, and Tomaz Prosen.
\newblock {Operator Entanglement in Local Quantum Circuits I: Chaotic
  Dual-Unitary Circuits}.
\newblock \emph{SciPost Phys.}, 8:\penalty0 067, 2020{\natexlab{b}}.
\newblock \doi{10.21468/SciPostPhys.8.4.067}.

\bibitem[Jamio{\l}kowski(1972)]{Jamiokowski1972L}
A.~Jamio{\l}kowski.
\newblock Linear transformations which preserve trace and positive
  semidefiniteness of operators.
\newblock \emph{Reports on Mathematical Physics}, 3\penalty0 (4):\penalty0
  275--278, 1972.
\newblock ISSN 0034-4877.
\newblock \doi{https://doi.org/10.1016/0034-4877(72)90011-0}.

\bibitem[Choi(1975)]{Choi1975C}
Man-Duen Choi.
\newblock Completely positive linear maps on complex matrices.
\newblock \emph{Linear Algebra and its Applications}, 10\penalty0 (3):\penalty0
  285--290, 1975.
\newblock ISSN 0024-3795.
\newblock \doi{https://doi.org/10.1016/0024-3795(75)90075-0}.

\bibitem[Bertini et~al.(2019)Bertini, Kos, and Prosen]{bertiniExact2019}
Bruno Bertini, Pavel Kos, and Toma\ifmmode \check{z}\else~\v{z}\fi{} Prosen.
\newblock Exact correlation functions for dual-unitary lattice models in $1+1$
  dimensions.
\newblock \emph{Phys. Rev. Lett.}, 123:\penalty0 210601, Nov 2019.
\newblock \doi{10.1103/PhysRevLett.123.210601}.

\bibitem[{Akhtar} et~al.(2024){Akhtar}, {Anand}, {Marshall}, and
  {You}]{akhtar2024dual}
Ahmed~A. {Akhtar}, Namit {Anand}, Jeffrey {Marshall}, and Yi-Zhuang {You}.
\newblock {Dual-Unitary Classical Shadow Tomography}.
\newblock \emph{arXiv e-prints}, art. arXiv:2404.01068, April 2024.
\newblock \doi{10.48550/arXiv.2404.01068}.

\bibitem[Hamma et~al.(2012)Hamma, Santra, and Zanardi]{hammaQuantum2012}
Alioscia Hamma, Siddhartha Santra, and Paolo Zanardi.
\newblock Quantum entanglement in random physical states.
\newblock \emph{Phys. Rev. Lett.}, 109:\penalty0 040502, Jul 2012.
\newblock \doi{10.1103/PhysRevLett.109.040502}.

\bibitem[{Hayden} et~al.(2016){Hayden}, {Nezami}, {Qi}, {Thomas}, {Walter}, and
  {Yang}]{Hayden2016H1601.01694}
Patrick {Hayden}, Sepehr {Nezami}, Xiao-Liang {Qi}, Nathaniel {Thomas}, Michael
  {Walter}, and Zhao {Yang}.
\newblock {Holographic duality from random tensor networks}.
\newblock \emph{Journal of High Energy Physics}, 2016\penalty0 (11):\penalty0
  9, November 2016.
\newblock \doi{10.1007/JHEP11(2016)009}.

\bibitem[{Da Liao}(2023)]{liaoControllable2023}
Yuan {Da Liao}.
\newblock {Controllable Incremental Algorithm for Entanglement Entropy in
  Quantum Monte Carlo Simulations}.
\newblock \emph{arXiv e-prints}, art. arXiv:2307.10602, July 2023.
\newblock \doi{10.48550/arXiv.2307.10602}.

\bibitem[Zhang et~al.(2024)Zhang, Pan, Chen, Sun, and Meng]{zhangIntegral2024}
Xu~Zhang, Gaopei Pan, Bin-Bin Chen, Kai Sun, and Zi~Yang Meng.
\newblock Integral algorithm of exponential observables for interacting
  fermions in quantum monte carlo simulations.
\newblock \emph{Phys. Rev. B}, 109:\penalty0 205147, May 2024.
\newblock \doi{10.1103/PhysRevB.109.205147}.

\bibitem[Zhou et~al.(2024)Zhou, Meng, Qi, and Da~Liao]{zhouIncremental2024}
Xuan Zhou, Zi~Yang Meng, Yang Qi, and Yuan Da~Liao.
\newblock Incremental swap operator for entanglement entropy: Application for
  exponential observables in quantum monte carlo simulation.
\newblock \emph{Phys. Rev. B}, 109:\penalty0 165106, Apr 2024.
\newblock \doi{10.1103/PhysRevB.109.165106}.

\bibitem[Hastings et~al.(2010)Hastings, Gonz\'alez, Kallin, and
  Melko]{hastingsMeasuring2010}
Matthew~B. Hastings, Iv\'an Gonz\'alez, Ann~B. Kallin, and Roger~G. Melko.
\newblock Measuring renyi entanglement entropy in quantum monte carlo
  simulations.
\newblock \emph{Phys. Rev. Lett.}, 104:\penalty0 157201, Apr 2010.
\newblock \doi{10.1103/PhysRevLett.104.157201}.

\bibitem[Luitz et~al.(2014)Luitz, Plat, Laflorencie, and
  Alet]{improvingLuitz2014}
David~J. Luitz, Xavier Plat, Nicolas Laflorencie, and Fabien Alet.
\newblock Improving entanglement and thermodynamic r\'enyi entropy measurements
  in quantum monte carlo.
\newblock \emph{Phys. Rev. B}, 90:\penalty0 125105, Sep 2014.
\newblock \doi{10.1103/PhysRevB.90.125105}.

\bibitem[Song et~al.(2024{\natexlab{a}})Song, Wang, and
  Meng]{song2024resummation}
Menghan Song, Ting-Tung Wang, and Zi~Yang Meng.
\newblock Resummation-based quantum monte carlo for entanglement entropy
  computation.
\newblock \emph{Phys. Rev. B}, 110:\penalty0 115117, Sep 2024{\natexlab{a}}.
\newblock \doi{10.1103/PhysRevB.110.115117}.

\bibitem[D'Emidio(2020)]{demidioEntanglement2020}
Jonathan D'Emidio.
\newblock Entanglement entropy from nonequilibrium work.
\newblock \emph{Phys. Rev. Lett.}, 124:\penalty0 110602, Mar 2020.
\newblock \doi{10.1103/PhysRevLett.124.110602}.

\bibitem[Zhao et~al.(2022{\natexlab{a}})Zhao, Wang, Yan, Cheng, and
  Meng]{zhaoScaling2022}
Jiarui Zhao, Yan-Cheng Wang, Zheng Yan, Meng Cheng, and Zi~Yang Meng.
\newblock Scaling of entanglement entropy at deconfined quantum criticality.
\newblock \emph{Phys. Rev. Lett.}, 128:\penalty0 010601, Jan
  2022{\natexlab{a}}.
\newblock \doi{10.1103/PhysRevLett.128.010601}.

\bibitem[Zhao et~al.(2022{\natexlab{b}})Zhao, Chen, Wang, Yan, Cheng, and
  Meng]{zhaoMeasuring2022}
Jiarui Zhao, Bin-Bin Chen, Yan-Cheng Wang, Zheng Yan, Meng Cheng, and Zi~Yang
  Meng.
\newblock Measuring r{\'e}nyi entanglement entropy with high efficiency and
  precision in quantum monte carlo simulations.
\newblock \emph{npj Quantum Materials}, 7:\penalty0 69, 2022{\natexlab{b}}.
\newblock \doi{10.1038/s41535-022-00476-0}.

\bibitem[Song et~al.(2024{\natexlab{b}})Song, Zhao, Meng, Xu, and
  Cheng]{song2024Extracting}
Menghan Song, Jiarui Zhao, Zi~Yang Meng, Cenke Xu, and Meng Cheng.
\newblock {Extracting subleading corrections in entanglement entropy at quantum
  phase transitions}.
\newblock \emph{SciPost Phys.}, 17:\penalty0 010, 2024{\natexlab{b}}.
\newblock \doi{10.21468/SciPostPhys.17.1.010}.

\bibitem[{Wang} et~al.(2024){Wang}, {Song}, {Meng}, and
  {Grover}]{wangAnalog2024}
Ting-Tung {Wang}, Menghan {Song}, Zi~Yang {Meng}, and Tarun {Grover}.
\newblock {An analog of topological entanglement entropy for mixed states}.
\newblock \emph{arXiv e-prints}, art. arXiv:2407.20500, July 2024.
\newblock \doi{10.48550/arXiv.2407.20500}.

\bibitem[Song et~al.(2025)Song, Zhao, Cheng, Xu, Scherer, Janssen, and
  Meng]{songEvolution2025}
Menghan Song, Jiarui Zhao, Meng Cheng, Cenke Xu, Michael Scherer, Lukas
  Janssen, and Zi~Yang Meng.
\newblock Evolution of entanglement entropy at su(n) deconfined quantum
  critical points.
\newblock \emph{Science Advances}, 11\penalty0 (6):\penalty0 eadr0634, 2025.
\newblock \doi{10.1126/sciadv.adr0634}.

\bibitem[Wang et~al.(2025)Wang, Song, Lyu, Witczak-Krempa, and
  Meng]{wangEntanglement2025}
Ting-Tung Wang, Menghan Song, Liuke Lyu, William Witczak-Krempa, and Zi~Yang
  Meng.
\newblock Entanglement microscopy and tomography in many-body systems.
\newblock \emph{Nature Communications}, 16:\penalty0 96, 2025.
\newblock \doi{10.1038/s41467-024-55354-z}.

\bibitem[git()]{github}
The code can be found on \url{https://github.com/songmengh/dual_unitary_MC}.

\bibitem[{Ippoliti} et~al.(2023){Ippoliti}, {Li}, {Rakovszky}, and
  {Khemani}]{Ippoliti2023O2212.11963}
Matteo {Ippoliti}, Yaodong {Li}, Tibor {Rakovszky}, and Vedika {Khemani}.
\newblock {Operator Relaxation and the Optimal Depth of Classical Shadows}.
\newblock \emph{Physical Review Letters}, 130\penalty0 (23):\penalty0 230403,
  June 2023.
\newblock \doi{10.1103/PhysRevLett.130.230403}.

\bibitem[{Huang} et~al.(2020){Huang}, {Kueng}, and
  {Preskill}]{Huang2020P2002.08953}
Hsin-Yuan {Huang}, Richard {Kueng}, and John {Preskill}.
\newblock {Predicting many properties of a quantum system from very few
  measurements}.
\newblock \emph{Nature Physics}, 16\penalty0 (10):\penalty0 1050--1057, June
  2020.
\newblock \doi{10.1038/s41567-020-0932-7}.

\bibitem[{Ohliger} et~al.(2013){Ohliger}, {Nesme}, and
  {Eisert}]{Ohliger2013E1204.5735}
M.~{Ohliger}, V.~{Nesme}, and J.~{Eisert}.
\newblock {Efficient and feasible state tomography of quantum many-body
  systems}.
\newblock \emph{New Journal of Physics}, 15\penalty0 (1):\penalty0 015024,
  January 2013.
\newblock \doi{10.1088/1367-2630/15/1/015024}.

\bibitem[{Guta} et~al.(2020){Guta}, {Kahn}, {Kueng}, and
  {Tropp}]{Guta2020F1809.11162}
Madalin {Guta}, Jonas {Kahn}, Richard {Kueng}, and Joel~A. {Tropp}.
\newblock {Fast state tomography with optimal error bounds}.
\newblock \emph{Journal of Physics A: Mathematical and Theoretical},
  53\penalty0 (20):\penalty0 204001, April 2020.
\newblock \doi{10.1088/1751-8121/ab8111}.

\bibitem[{Elben} et~al.(2020){Elben}, {Kueng}, {Huang}, {van Bijnen}, {Kokail},
  {Dalmonte}, {Calabrese}, {Kraus}, {Preskill}, {Zoller}, and
  {Vermersch}]{Elben2020M2007.06305}
Andreas {Elben}, Richard {Kueng}, Hsin-Yuan~Robert {Huang}, Rick {van Bijnen},
  Christian {Kokail}, Marcello {Dalmonte}, Pasquale {Calabrese}, Barbara
  {Kraus}, John {Preskill}, Peter {Zoller}, and Beno{\^\i}t {Vermersch}.
\newblock {Mixed-State Entanglement from Local Randomized Measurements}.
\newblock \emph{Physical Review Letters}, 125\penalty0 (20):\penalty0 200501,
  November 2020.
\newblock \doi{10.1103/PhysRevLett.125.200501}.

\bibitem[{Huang} et~al.(2021{\natexlab{a}}){Huang}, {Broughton}, {Mohseni},
  {Babbush}, {Boixo}, {Neven}, and {McClean}]{Huang2021P2011.01938}
Hsin-Yuan {Huang}, Michael {Broughton}, Masoud {Mohseni}, Ryan {Babbush},
  Sergio {Boixo}, Hartmut {Neven}, and Jarrod~R. {McClean}.
\newblock {Power of data in quantum machine learning}.
\newblock \emph{Nature Communications}, 12:\penalty0 2631, January
  2021{\natexlab{a}}.
\newblock \doi{10.1038/s41467-021-22539-9}.

\bibitem[{Huang} et~al.(2021{\natexlab{b}}){Huang}, {Kueng}, and
  {Preskill}]{Huang2021I2101.02464}
Hsin-Yuan {Huang}, Richard {Kueng}, and John {Preskill}.
\newblock {Information-Theoretic Bounds on Quantum Advantage in Machine
  Learning}.
\newblock \emph{Physical Review Letters}, 126\penalty0 (19):\penalty0 190505,
  May 2021{\natexlab{b}}.
\newblock \doi{10.1103/PhysRevLett.126.190505}.

\bibitem[{Huang} et~al.(2021{\natexlab{c}}){Huang}, {Kueng}, {Torlai},
  {Albert}, and {Preskill}]{Huang2021P2106.12627}
Hsin-Yuan {Huang}, Richard {Kueng}, Giacomo {Torlai}, Victor~V. {Albert}, and
  John {Preskill}.
\newblock {Provably efficient machine learning for quantum many-body problems}.
\newblock \emph{arXiv e-prints}, art. arXiv:2106.12627, June
  2021{\natexlab{c}}.
\newblock \doi{10.48550/arXiv.2106.12627}.

\bibitem[{Huang} et~al.(2021{\natexlab{d}}){Huang}, {Kueng}, and
  {Preskill}]{Huang2021E2103.07510}
Hsin-Yuan {Huang}, Richard {Kueng}, and John {Preskill}.
\newblock {Efficient Estimation of Pauli Observables by Derandomization}.
\newblock \emph{Physical Review Letters}, 127\penalty0 (3):\penalty0 030503,
  July 2021{\natexlab{d}}.
\newblock \doi{10.1103/PhysRevLett.127.030503}.

\bibitem[Levy et~al.(2024)Levy, Luo, and Clark]{Levy2021C2110.02965}
Ryan Levy, Di~Luo, and Bryan~K. Clark.
\newblock {Classical shadows for quantum process tomography on near-term
  quantum computers}.
\newblock \emph{Phys. Rev. Res.}, 6\penalty0 (1):\penalty0 013029, 2024.
\newblock \doi{10.1103/PhysRevResearch.6.013029}.

\bibitem[{Acharya} et~al.(2021){Acharya}, {Saha}, and
  {Sengupta}]{Acharya2021S2105.05992}
Atithi {Acharya}, Siddhartha {Saha}, and Anirvan~M. {Sengupta}.
\newblock {Shadow tomography based on informationally complete positive
  operator-valued measure}.
\newblock \emph{{Phys. Rev. A}}, {104}\penalty0 ({5}):\penalty0 {052418},
  November 2021.
\newblock \doi{10.1103/PhysRevA.104.052418}.

\bibitem[{Hu} and {You}(2022)]{Hu2022H2102.10132}
Hong-Ye {Hu} and Yi-Zhuang {You}.
\newblock {Hamiltonian-driven shadow tomography of quantum states}.
\newblock \emph{Physical Review Research}, 4\penalty0 (1):\penalty0 013054,
  January 2022.
\newblock \doi{10.1103/PhysRevResearch.4.013054}.

\bibitem[{Bu} et~al.(2022){Bu}, {Enshan Koh}, {Garcia}, and
  {Jaffe}]{Bu2022C2202.03272}
Kaifeng {Bu}, Dax {Enshan Koh}, Roy~J. {Garcia}, and Arthur {Jaffe}.
\newblock {Classical shadows with Pauli-invariant unitary ensembles}.
\newblock \emph{arXiv e-prints}, art. arXiv:2202.03272, February 2022.
\newblock \doi{10.48550/arXiv.2202.03272}.

\bibitem[Seif et~al.(2023)Seif, Cian, Zhou, Chen, and
  Jiang]{Seif2022S2203.07309}
Alireza Seif, Ze-Pei Cian, Sisi Zhou, Senrui Chen, and Liang Jiang.
\newblock {Shadow Distillation: Quantum Error Mitigation with Classical Shadows
  for Near-Term Quantum Processors}.
\newblock \emph{PRX Quantum}, 4\penalty0 (1):\penalty0 010303, 2023.
\newblock \doi{10.1103/PRXQuantum.4.010303}.

\bibitem[{Hu} et~al.(2022){Hu}, {LaRose}, {You}, {Rieffel}, and
  {Wang}]{Hu2022L2203.07263}
Hong-Ye {Hu}, Ryan {LaRose}, Yi-Zhuang {You}, Eleanor {Rieffel}, and Zhihui
  {Wang}.
\newblock {Logical shadow tomography: Efficient estimation of error-mitigated
  observables}.
\newblock \emph{arXiv e-prints}, art. arXiv:2203.07263, March 2022.
\newblock \doi{10.48550/arXiv.2203.07263}.

\bibitem[{Chau Nguyen} et~al.(2022){Chau Nguyen}, {Lennart B{\"o}nsel},
  {Steinberg}, and {G{\"u}hne}]{Chau-Nguyen2022O2205.08990}
H.~{Chau Nguyen}, Jan {Lennart B{\"o}nsel}, Jonathan {Steinberg}, and Otfried
  {G{\"u}hne}.
\newblock {Optimising shadow tomography with generalised measurements}.
\newblock \emph{arXiv e-prints}, art. arXiv:2205.08990, May 2022.
\newblock \doi{10.48550/arXiv.2205.08990}.

\bibitem[{Hadfield} et~al.(2020){Hadfield}, {Bravyi}, {Raymond}, and
  {Mezzacapo}]{Hadfield2022M2006.15788}
Charles {Hadfield}, Sergey {Bravyi}, Rudy {Raymond}, and Antonio {Mezzacapo}.
\newblock {Measurements of Quantum Hamiltonians with Locally-Biased Classical
  Shadows}.
\newblock \emph{arXiv e-prints}, art. arXiv:2006.15788, June 2020.
\newblock \doi{10.48550/arXiv.2006.15788}.

\bibitem[{Huang} et~al.(2022){Huang}, {Broughton}, {Cotler}, {Chen}, {Li},
  {Mohseni}, {Neven}, {Babbush}, {Kueng}, {Preskill}, and
  {McClean}]{Huang2022Q2112.00778}
Hsin-Yuan {Huang}, Michael {Broughton}, Jordan {Cotler}, Sitan {Chen}, Jerry
  {Li}, Masoud {Mohseni}, Hartmut {Neven}, Ryan {Babbush}, Richard {Kueng},
  John {Preskill}, and Jarrod~R. {McClean}.
\newblock {Quantum advantage in learning from experiments}.
\newblock \emph{Science}, 376\penalty0 (6598):\penalty0 1182--1186, June 2022.
\newblock \doi{10.1126/science.abn7293}.

\bibitem[{Enshan Koh} and {Grewal}(2022)]{Enshan-Koh2022C2011.11580}
Dax {Enshan Koh} and Sabee {Grewal}.
\newblock {Classical Shadows With Noise}.
\newblock \emph{{Quantum}}, 6:\penalty0 arXiv:2011.11580, August 2022.
\newblock \doi{10.48550/arXiv.2011.11580}.

\bibitem[{Hao Low}(2022)]{Hao-Low2022C2208.08964}
Guang {Hao Low}.
\newblock {Classical shadows of fermions with particle number symmetry}.
\newblock \emph{arXiv e-prints}, art. arXiv:2208.08964, August 2022.
\newblock \doi{10.48550/arXiv.2208.08964}.

\bibitem[Bertoni et~al.(2024)Bertoni, Haferkamp, Hinsche, Ioannou, Eisert, and
  Pashayan]{Bertoni2022S2209.12924}
Christian Bertoni, Jonas Haferkamp, Marcel Hinsche, Marios Ioannou, Jens
  Eisert, and Hakop Pashayan.
\newblock Shallow shadows: Expectation estimation using low-depth random
  clifford circuits.
\newblock \emph{Phys. Rev. Lett.}, 133:\penalty0 020602, Jul 2024.
\newblock \doi{10.1103/PhysRevLett.133.020602}.
\newblock URL \url{https://link.aps.org/doi/10.1103/PhysRevLett.133.020602}.

\bibitem[{Arienzo} et~al.(2022){Arienzo}, {Heinrich}, {Roth}, and
  {Kliesch}]{Arienzo2022C2211.09835}
Mirko {Arienzo}, Markus {Heinrich}, Ingo {Roth}, and Martin {Kliesch}.
\newblock {Closed-form analytic expressions for shadow estimation with
  brickwork circuits}.
\newblock \emph{arXiv e-prints}, art. arXiv:2211.09835, November 2022.
\newblock \doi{10.48550/arXiv.2211.09835}.

\bibitem[{Hu} et~al.(2023){Hu}, {Choi}, and {You}]{Hu2023C2107.04817}
Hong-Ye {Hu}, Soonwon {Choi}, and Yi-Zhuang {You}.
\newblock {Classical Shadow Tomography with Locally Scrambled Quantum
  Dynamics}.
\newblock \emph{Physical Review Research}, 5\penalty0 (2):\penalty0
  arXiv:2107.04817, April 2023.
\newblock \doi{10.1103/PhysRevResearch.5.023027}.

\bibitem[{Akhtar} et~al.(2023){Akhtar}, {Hu}, and {You}]{Akhtar2023S2209.02093}
Ahmed~A. {Akhtar}, Hong-Ye {Hu}, and Yi-Zhuang {You}.
\newblock {Scalable and Flexible Classical Shadow Tomography with Tensor
  Networks}.
\newblock \emph{Quantum}, 7:\penalty0 1026, June 2023.
\newblock \doi{10.22331/q-2023-06-01-1026}.

\bibitem[{Zhang} et~al.(2024){Zhang}, {Feng}, {Ippoliti}, and
  {You}]{zhangHolographic2024}
Shuhan {Zhang}, Xiaozhou {Feng}, Matteo {Ippoliti}, and Yi-Zhuang {You}.
\newblock {Holographic Classical Shadow Tomography}.
\newblock \emph{arXiv e-prints}, art. arXiv:2406.11788, June 2024.
\newblock \doi{10.48550/arXiv.2406.11788}.

\bibitem[Akhtar et~al.(2024)Akhtar, Hu, and You]{Akhtar2024M2308.01653}
Ahmed~A. Akhtar, Hong-Ye Hu, and Yi-Zhuang You.
\newblock Measurement-induced criticality is tomographically optimal.
\newblock \emph{Phys. Rev. B}, 109:\penalty0 094209, Mar 2024.
\newblock \doi{10.1103/PhysRevB.109.094209}.

\bibitem[{Hu} et~al.(2024){Hu}, {Gu}, {Majumder}, {Ren}, {Zhang}, {Wang},
  {You}, {Minev}, {Yelin}, and {Seif}]{Hu2024D2402.17911}
Hong-Ye {Hu}, Andi {Gu}, Swarnadeep {Majumder}, Hang {Ren}, Yipei {Zhang},
  Derek~S. {Wang}, Yi-Zhuang {You}, Zlatko {Minev}, Susanne~F. {Yelin}, and
  Alireza {Seif}.
\newblock {Demonstration of Robust and Efficient Quantum Property Learning with
  Shallow Shadows}.
\newblock \emph{arXiv e-prints}, art. arXiv:2402.17911, February 2024.
\newblock \doi{10.48550/arXiv.2402.17911}.

\bibitem[{Liu} et~al.(2023{\natexlab{b}}){Liu}, {Hao}, and
  {Hu}]{Liu2023P2311.00695}
Zhenhuan {Liu}, Zihan {Hao}, and Hong-Ye {Hu}.
\newblock {Predicting Arbitrary State Properties from Single Hamiltonian Quench
  Dynamics}.
\newblock \emph{arXiv e-prints}, art. arXiv:2311.00695, November
  2023{\natexlab{b}}.
\newblock \doi{10.48550/arXiv.2311.00695}.

\bibitem[{Shen} et~al.(2024){Shen}, {Buzali}, {Hu}, {Klymko}, {Camps}, {Yelin},
  and {Van Beeumen}]{Shen2024E2409.13691}
Yizhi {Shen}, Alex {Buzali}, Hong-Ye {Hu}, Katherine {Klymko}, Daan {Camps},
  Susanne~F. {Yelin}, and Roel {Van Beeumen}.
\newblock {Efficient Measurement-Driven Eigenenergy Estimation with Classical
  Shadows}.
\newblock \emph{arXiv e-prints}, art. arXiv:2409.13691, September 2024.
\newblock \doi{10.48550/arXiv.2409.13691}.

\bibitem[Ippoliti and Khemani(2024)]{Ippoliti2024L}
Matteo Ippoliti and Vedika Khemani.
\newblock Learnability transitions in monitored quantum dynamics via
  eavesdropper's classical shadows.
\newblock \emph{PRX Quantum}, 5:\penalty0 020304, Apr 2024.
\newblock \doi{10.1103/PRXQuantum.5.020304}.

\bibitem[{Ippoliti}(2024)]{Ippoliti2024C2305.10723}
Matteo {Ippoliti}.
\newblock {Classical shadows based on locally-entangled measurements}.
\newblock \emph{Quantum}, 8:\penalty0 1293, March 2024.
\newblock \doi{10.22331/q-2024-03-21-1293}.

\bibitem[{Chen} et~al.(2024){Chen}, {Gong}, and {Ye}]{Chen2024O2404.19105}
Sitan {Chen}, Weiyuan {Gong}, and Qi~{Ye}.
\newblock {Optimal tradeoffs for estimating Pauli observables}.
\newblock \emph{arXiv e-prints}, art. arXiv:2404.19105, April 2024.
\newblock \doi{10.48550/arXiv.2404.19105}.

\bibitem[{Rakovszky} et~al.(2018){Rakovszky}, {Pollmann}, and {von
  Keyserlingk}]{Rakovszky2018D1710.09827}
Tibor {Rakovszky}, Frank {Pollmann}, and C.~W. {von Keyserlingk}.
\newblock {Diffusive Hydrodynamics of Out-of-Time-Ordered Correlators with
  Charge Conservation}.
\newblock \emph{Physical Review X}, 8\penalty0 (3):\penalty0 031058, July 2018.
\newblock \doi{10.1103/PhysRevX.8.031058}.

\bibitem[{Hunter-Jones}(2018)]{Hunter-Jones2018O1812.08219}
Nicholas {Hunter-Jones}.
\newblock {Operator growth in random quantum circuits with symmetry}.
\newblock \emph{arXiv e-prints}, art. arXiv:1812.08219, December 2018.
\newblock \doi{10.48550/arXiv.1812.08219}.

\bibitem[Agrawal et~al.(2022)Agrawal, Zabalo, Chen, Wilson, Potter, Pixley,
  Gopalakrishnan, and Vasseur]{agrawalEntanglement2022}
Utkarsh Agrawal, Aidan Zabalo, Kun Chen, Justin~H. Wilson, Andrew~C. Potter,
  J.~H. Pixley, Sarang Gopalakrishnan, and Romain Vasseur.
\newblock Entanglement and charge-sharpening transitions in u(1) symmetric
  monitored quantum circuits.
\newblock \emph{Phys. Rev. X}, 12:\penalty0 041002, Oct 2022.
\newblock \doi{10.1103/PhysRevX.12.041002}.

\bibitem[{Foligno} et~al.(2024){Foligno}, {Calabrese}, and
  {Bertini}]{folignoNonequilibrium2024}
Alessandro {Foligno}, Pasquale {Calabrese}, and Bruno {Bertini}.
\newblock {Non-equilibrium dynamics of charged dual-unitary circuits}.
\newblock \emph{arXiv e-prints}, art. arXiv:2407.21786, July 2024.
\newblock \doi{10.48550/arXiv.2407.21786}.

\bibitem[{Chang} et~al.(2024){Chang}, {Shrotriya}, {Ho}, and
  {Ippoliti}]{Chang2024D2408.15325}
Rui-An {Chang}, Harshank {Shrotriya}, Wen~Wei {Ho}, and Matteo {Ippoliti}.
\newblock {Deep thermalization under charge-conserving quantum dynamics}.
\newblock \emph{arXiv e-prints}, art. arXiv:2408.15325, August 2024.
\newblock \doi{10.48550/arXiv.2408.15325}.

\bibitem[{Fisher} et~al.(2023){Fisher}, {Khemani}, {Nahum}, and
  {Vijay}]{Fisher2023R2207.14280}
Matthew P.~A. {Fisher}, Vedika {Khemani}, Adam {Nahum}, and Sagar {Vijay}.
\newblock {Random Quantum Circuits}.
\newblock \emph{Annual Review of Condensed Matter Physics}, 14:\penalty0
  335--379, March 2023.
\newblock \doi{10.1146/annurev-conmatphys-031720-030658}.

\end{thebibliography}

\end{document}